\documentclass[aps,prl,twocolumn,showpacs,superscriptaddress]{revtex4-2}
\usepackage{amsmath}
\usepackage{amssymb}   
\usepackage{amsthm}   
\usepackage{bm}
\usepackage{float}
\usepackage{url}
\usepackage{graphicx}
\usepackage{hyperref}
\hypersetup{
  colorlinks   = true, %Colours links instead of ugly boxes
  urlcolor     = blue, %Colour for external hyperlinks
  linkcolor    = black, %Colour of internal links
  citecolor   = black %Colour of citations
}

\newcommand{\logfn}[2]{\left({#1} - {#2}\right) \log \left( \frac{ {#1}}{{#2}} \right)}
\newcommand{\logfnInline}[2]{\left({#1} - {#2}\right) \log \left(  {#1}/{#2} \right)}
\renewcommand{\d}{\text{d}}

\newcommand{\be}{\begin{equation}}
\newcommand{\ee}{\end{equation}}
\newcommand{\ba}{\begin{align}}
\newcommand{\ea}{\end{align}}

\begin{document}
\raggedbottom
\title{Estimating entropy production from waiting time distributions}

\author{Dominic J. Skinner} 
\affiliation{Department of Mathematics, Massachusetts Institute of Technology, Cambridge Massachusetts 02139-4307, USA}
\author{J\"{o}rn Dunkel}
\affiliation{Department of Mathematics, Massachusetts Institute of Technology, Cambridge Massachusetts 02139-4307, USA}
\date{\today}

\begin{abstract}
Living systems operate far from thermal equilibrium by converting the chemical potential of ATP  into mechanical work to achieve growth, replication or locomotion. Given time series observations of intra-, inter- or multicellular processes, a key challenge is to detect non-equilibrium behavior and quantify the rate of free energy consumption. Obtaining reliable bounds on energy consumption and entropy production directly from experimental data remains difficult in practice as many degrees of freedom typically are hidden to the observer, so that the accessible coarse-grained dynamics may not obviously violate detailed balance. Here, we introduce a novel method for bounding the entropy production of physical and living systems which uses only the waiting time statistics of hidden Markov processes and hence can be directly applied to experimental data. By determining a universal limiting curve, we infer entropy production bounds from experimental data for gene regulatory networks, mammalian behavioral dynamics and numerous other biological processes. Further considering the asymptotic limit of increasingly precise biological timers, we estimate the necessary entropic cost of heartbeat regulation in humans, dogs and mice.
\end{abstract}

\maketitle
Living systems break time reversal symmetry~\cite{Parrondo_2009,England_2013}. 
Growth~\cite{WANG2010}, self-replication~\cite{England_2013} and locomotion~\cite{Nirody_2017} are all irreversible processes, violating the principle of detailed balance, which states that every forward microstate trajectory is as likely to occur as its time reversed counterpart~\cite{Parrondo_2009,Seifert_2012,Maes_2002,Fakhri2016}.
Irreversibility places thermodynamic constraints on sensing~\cite{Lan2012,harvey2020universal}, reproduction~\cite{England_2013}, and signaling~\cite{Thurley2014}, and trade-offs between free energy consumption and the precision, speed, or accuracy of performing some function~\cite{Lan2012,Cao2015}.
To quantify these constraints, one has to bound the rate at which free energy is consumed, or equivalently the entropy production rate (EPR)~\cite{Seifert_2005,Seifert_2012,Seifert_AnnRev}. However, most experimental measurements can only observe a small number of degrees of freedom~\cite{Seifert_AnnRev,Esposito2012,Teza2020}, making such inference challenging. In particular, many systems are only accessible at the coarsest possible level of two states, such as a gene switching~\cite{Suter2011}, or a molecule docking and undocking to a sensor~\cite{Wingreen_2013,harvey2020universal}.
In these cases, commonly used inference tools cannot be applied, due to the absence of observable coarse-grained currents~\cite{TUR_proof_2016,Li_2019,Gingrich_natPhys,Esposito2017}, and as the forward and reverse trajectories need not be asymmetric~\cite{Gnesotto2020,Parrondo_2010,Horowitz_nocurrent, Julicher_arrow}, even if the underlying system operates far from equilibrium.

\par
Here, we introduce a broadly applicable method that makes it possible to estimate the EPR  of a coarse-grained two-state dynamics by measuring the variance of time spent in a state.  This is achieved by finding a canonical formulation and then solving a numerical optimization problem to obtain a universal limiting curve, above which the EPRs of all systems with the observed waiting time statistics must lie. We illustrate that this method outperforms a recently introduced thermodynamic uncertainty relation (TUR)~\cite{TUR_proof_2016,Gingrich_natPhys,harvey2020universal,Hasegawa_2020} on synthetic test data,  and demonstrate its practical usefulness in applications to experimental data for a gene regulatory network~\cite{Suter2011}, the behavioral dynamics of cows~\cite{TOLKAMP2010}, and several other biological processes (Table S1~\cite{SM}).  
Furthermore, by considering a stochastic timer~\cite{Seifert_2016,Horowitz_clock,Pearson2021,Junco2020}, we derive an asymptotic formula relating the waiting time variance and the EPR. This analytical result can be used to bound the EPR required to maintain precise biophysical timing processes, which we illustrate on data from experimental measurements~\cite{physioZoo,UMETANI1998} of heartbeats of humans, dogs and mice.

\par 
We start from the standard assumption~\cite{Seifert_2012} that meso-scale systems  in contact 
with a heat bath can be described by a Markovian stochastic dynamics on a finite set of discrete states $\{1,\dots, N_T\}$. 
Transitions from state $i$ to $j$ occur at rate $W_{ij}$, so a probability distribution
over the states, $p_i(t)$, evolves according to
\be\label{e:Master}
\frac{\d}{\d t} p_i = \sum_{j} p_j W_{ji},
\ee
with $W_{ii} = - \sum_{j \neq i} W_{ij}$~\cite{Seifert_2012,van1992stochastic}. Assuming that the system is irreducible, meaning there exists a path
with non-zero probability between any two states, Eq.~\eqref{e:Master} has a unique stationary
state $\pi=(\pi_i)$, satisfying $0 = \pi W$, and which every initial probability distribution
tends to~\cite{rubino_sericola_1989}. The EPR at steady state $\pi$ is~\cite{Gingrich_natPhys},
\be
\sigma = k_B \sum_{i < j} \logfn{\pi_i W_{ij}}{\pi_j W_{ji}},
\ee
which for isothermal systems, when multiplied by the bath temperature, is the rate of free energy dissipation required to maintain the state away from equilibrium~\cite{Qian_2013}.
The system is in thermal equilibrium when the EPR vanishes, which is 
exactly when detailed balance is satisfied, $\pi_i W_{ij} = \pi_j W_{ji}$,
for all $i$, $j$.
%%%%%%%%%%%%%%%% Fig 1
\begin{figure}\centering
\includegraphics[width=0.8\columnwidth]{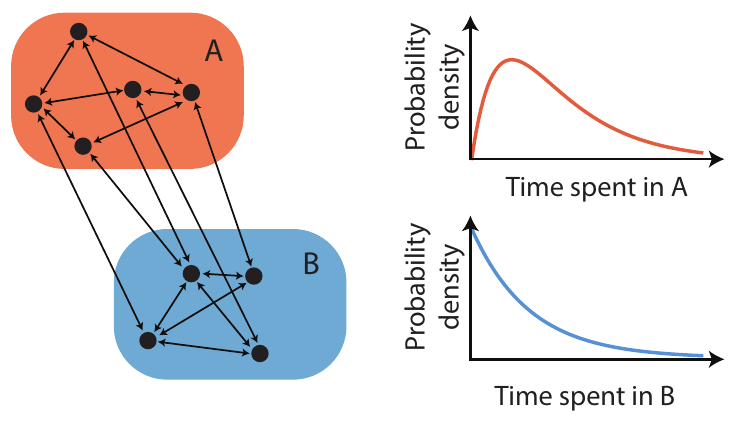}
\caption{\label{fig:TwoState}A Markovian process on discrete states (black circles), observed at a coarse-grained 
level, with only the metastates $A$ and $B$ visible to the observer (left). From observing this system we can
deduce the distribution of time spent in metastates $A$ and $B$ (right). A non-monotonic waiting time distribution signals~\cite{Tu_2008} out-of-equilibrium dynamics.}
\end{figure}
%%%%%%%%%%%%%%%% 

\par 
Only a coarse-grained view of the system is typically available in
experiments~\cite{Horowitz_nocurrent,Skinner2021}, with the observed dynamics taking place on a set of metastates, 
where a metastate $I$ may contain several underlying states $i$ (Fig.~\ref{fig:TwoState}).
Here, we focus on the common case where only two complementary metastates, $A$ and $B$, are accessible  and the observed trajectory jumps between them (Fig.~\ref{fig:TwoState}).  After observing many such jumps,
the distributions $f_A(t)$ and $f_B(t)$ of time spent in $A$ and $B$ can be empirically reconstructed (Fig.~\ref{fig:TwoState}).  As the coarse-grained observations are non-Markov~\cite{Jurgens2021}, there may be additional information in conditional statistics~\cite{Skinner2021,Teza2020}, but these are hard to measure experimentally and we do not 
consider them here.  At first glance, the distributions $f_A(t)$ and $f_B(t)$ alone may not appear to contain much information about the underlying system;  however,  for an equilibrium system both $f_A(t)$ and $f_B(t)$ must decay monotonically~\cite{Tu_2008}, implying that the schematic example in Fig.~\ref{fig:TwoState} reflects an out-of-equilibrium dynamics.

\par %\subsubsection{Optimization framework}
To quantify the extent to which an observed two-state system is out-of-equilibrium, we reformulate the problem of EPR estimation within an optimization framework, extending the approach introduced in Ref.~\cite{Skinner2021}. Calling the
underlying system $\mathcal{S}$, the EPR is $\sigma(\mathcal{S})$, and the observed
waiting time distributions are $f_A(t,\mathcal{S})$ and $f_B(t,\mathcal{S})$. If $\mathcal{R}$ describes some
other underlying system with the same observables, we cannot know which is the true system. However, we can
find a lower bound on the EPR by minimizing over all such $\mathcal{R}$,
\begin{align}
\nonumber \sigma(\mathcal{S}) \geq \min_{\mathcal{R}} \{ \sigma(\mathcal{R}) | &f_A(t,\mathcal{S}) = f_A(t,\mathcal{R}), \\
&f_B(t,\mathcal{S}) = f_B(t,\mathcal{R}) \},
\end{align}
which holds, since $\mathcal{S}$ is contained in the set on the right. By construction, this is the 
tightest possible lower bound given the observed distributions $f_A$ and $f_B$. Limited experimental data 
may prevent us from precisely measuring the full distribution, but
we can typically measure the first few moments $\langle t^n \rangle_A$ and $\langle t^n \rangle_B$. We therefore introduce
the estimators $\sigma^{(n)}_{T}$,
\begin{align}\label{eq:sig_n}
\nonumber
\sigma(\mathcal{S}) \geq \min_{\mathcal{R}} \{ &\sigma(\mathcal{R}) |\, \langle t^k \rangle_{I,\mathcal{R}} = 
\langle t^k \rangle_{I,\mathcal{S}}, \\
&\text{for } I=A,B, k = 1,\dots n \} \equiv \sigma^{(n)}_{T},
\end{align}
where $\sigma^{(n)}_{T} \leq \sigma^{(n+1)}_{T}$, since increasing $n$ decreases the size of the set over which
minimization is performed.
%%%%%%%%%%%%%%%% Fig 2
\begin{figure*}
\includegraphics{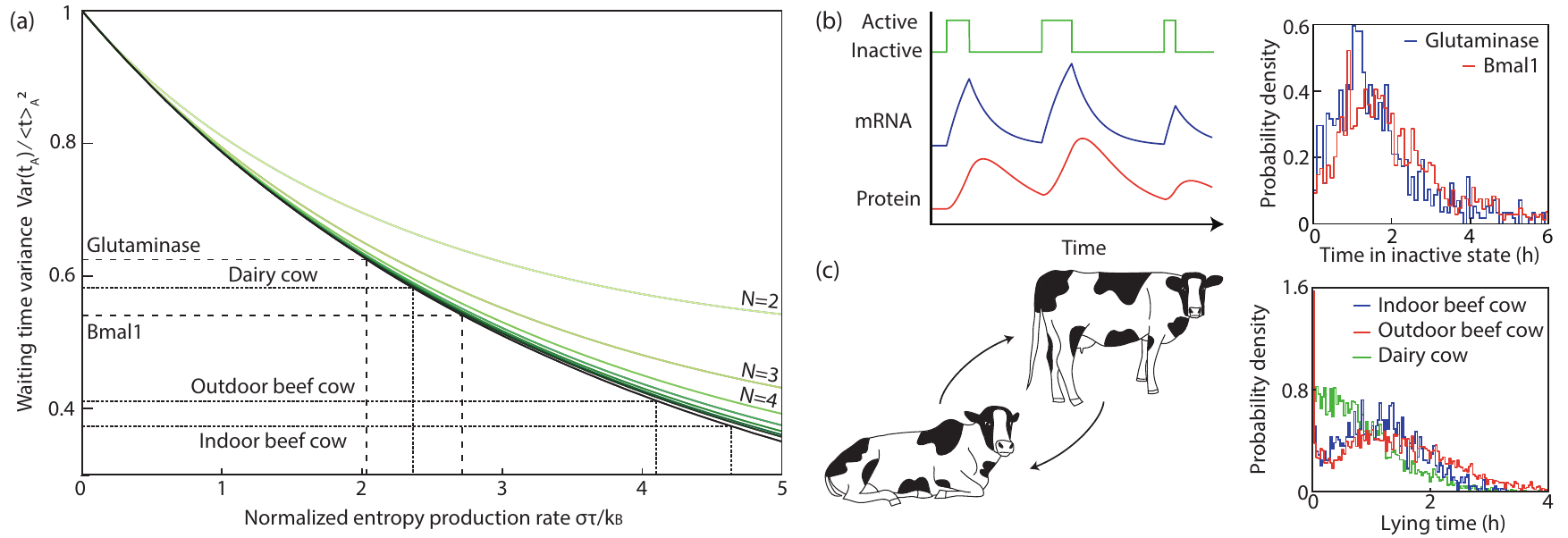}
\caption{\label{fig:TradeOff} Bounding the EPR from a single limiting curve. (a) All realizable systems lie above the curve $\Lambda$ (black) which represents the optimal variance-entropy production trade-off. Minimizing over increasing numbers of internal states leads to rapid convergence, with results shown for up to 8 internal states (green). Given measured values of the waiting time variance, a lower bound on entropy production can be read off, as demonstrated here for experimental data in (b, dashed) and (c, dotted). 
(b)  Genes stochastically switch from being
active to inactive, mRNA is produced when the gene is active, proteins are
produced when mRNA is present and both stochastically degrade (left). The distribution of time the gene spends
in the inactive state as measured in recent 
experiments, for the genes Glutaminase and Bmal1 (right, see also Fig. 2C in Ref.~\cite{Suter2011}).
(c) Distribution of time cows spent lying before standing across three experiments from
Ref.~\cite{TOLKAMP2010}.}
\end{figure*}
%%%%%%%%%%%%%%%% 
\par
Instead of specifying the transition rates of a system~$\mathcal{R}$, we can alternatively specify the net transition rates, $n_{ij} = \pi_i W_{ij}$, for $i\neq j$, provided they
satisfy mass conservation at each vertex, $\sum_j n_{ij} = \sum_j n_{ji}$, and independently specify the stationary state,~$\pi_i$. With $n_{ij}$ fixed, modifying $\pi_i$ does not affect the EPR. In particular, suppose that we have found the optimal $\pi_i$, $n_{ij}$,
subject to the constraints, then the fraction of time spent in $A$ is $r_A = \sum_{i \in A} \pi_i$, and similarly $r_B = \sum_{i\in B} \pi_i$. Rescaling $\pi_i \mapsto \pi_i / 2 r_A$ for $i\in A$ and $\pi_i \mapsto \pi_i / 2 r_B$ for $i\in B$,  does not affect the EPR, but makes the fraction of time spent in each metastate exactly~$1/2$.
The statistics are rescaled as $\langle t^k \rangle_I \mapsto (1/2r_I)^k \langle t^k \rangle_I$ for $I=A,B$.
The average time spent in $A$ or $B$ in the rescaled system is $\tau = (\langle t \rangle_A 
+ \langle t \rangle_B)/2$. A second rescaling $n_{ij} \mapsto \tau n_{ij}$ with $\pi$ fixed, changes EPR as $\sigma \mapsto \tau \sigma$, and the moments as $\langle t^k \rangle_I \mapsto
\langle t^k \rangle_I / \tau^k$. Therefore, we can rewrite Eq.~\eqref{eq:sig_n} as
\begin{align}
\sigma(\mathcal{S}) \geq \frac{1}{\tau} \min_{\mathcal{R}} \{ &\sigma | \langle t^k \rangle_{I, \mathcal{R}} =
\langle t^k \rangle_{I,\mathcal{S}} / \langle t \rangle_{I,\mathcal{S}}^k, \\
\nonumber &\text{ for } I=A,B,\ k = 2,\dots, n \},
\end{align} 
allowing us to optimize over
a single canonical system and rescale to bound any other non-canonical system.

\par
After the canonical rescaling, to express constraints in terms of the transition rates, we label the states so that the first $N$ belong to the metastate $A$, with $1 \leq N < N_T$, and define $(W_A)_{ij} = W_{i,j}$ for $i,j \leq N$, so $W_A$ represents the transitions within $A$. We also
write $\pi = (\pi_A, \pi_B)$, with $\pi_A$ the first $N$ components of the stationary distribution. 
For any such Markovian process in the stationary distribution, we have that $f_A(t) =  \pi_A W_A^2 \exp(W_A t) \bm{1}^T / (- \pi_A W_A \bm{1}^T)$~\cite{SM,rubino_sericola_1989}, 
and similar for $f_B(t)$. For the canonical system,  $\langle t \rangle_{A,\mathcal{R}} = 1 $, so $- \pi_A W_A \bm{1}^T = 1/2$, and higher moment constraints become
\be 
\langle t^k \rangle_{A,\mathcal{R}} = 2(-1)^{k+1} k! \pi_A W_A^{1-k} \bm{1}^T ,
\ee
and similar for $B$. While this provides a general minimization framework, from now on we will only impose the constraint on $\langle t^2 \rangle_{A,\mathcal{R}}$, 
which allows us to minimize over systems with $B$ consisting of a single state~\cite{SM}, and results in a single curve. Specifically, we minimize over $\pi_i >0$, $n_{ij} \geq 0$, subject to the linear constraints
$n_{ii} = -\sum_{j\neq i } n_{ij}$, $\sum_{j\neq i} n_{ij} = \sum_{j\neq i} n_{ji}$, together with $\langle t \rangle_{A,\mathcal{R}} = \langle t \rangle_{B,\mathcal{R}} = 1$,
and $\langle t^2 \rangle_{A,\mathcal{R}}$ fixed, finding a curve $k_B \Lambda( \langle t^2 \rangle_{A,\mathcal{R}})$
in the $\sigma$--$\langle t^2 \rangle_{A,\mathcal{R}}$ plane, or equivalently the $\sigma$--$\text{Var} \, t_{A,\mathcal{R}}$ plane, above which all Markovian systems must lie [Fig.~\ref{fig:TradeOff}(a)].  For arbitrary systems, this bound becomes
\be
\sigma_T = \left[ \frac{2k_B}{ \langle t \rangle_A + \langle t \rangle_B }\right] \Lambda \left( \frac{ \text{Var}\, t_A}{
\langle t \rangle_A^2} \right),
\label{e:sig_T}
\ee
which we use throughout to bound the EPR. The function $\Lambda(x)$ can be computed numerically~\cite{SM}; it converges rapidly as the number $N$ of internal states of $A$ is increased [Fig.~\ref{fig:TradeOff}(a)]. Given the tabulated values of $\Lambda$, we can apply Eq.~\eqref{e:sig_T} directly to experimental data. Before doing so, it is instructive to demonstrate the usefulness of $\sigma_T$ in simulations of an active sensor.

\par
\begin{figure}\centering
\includegraphics{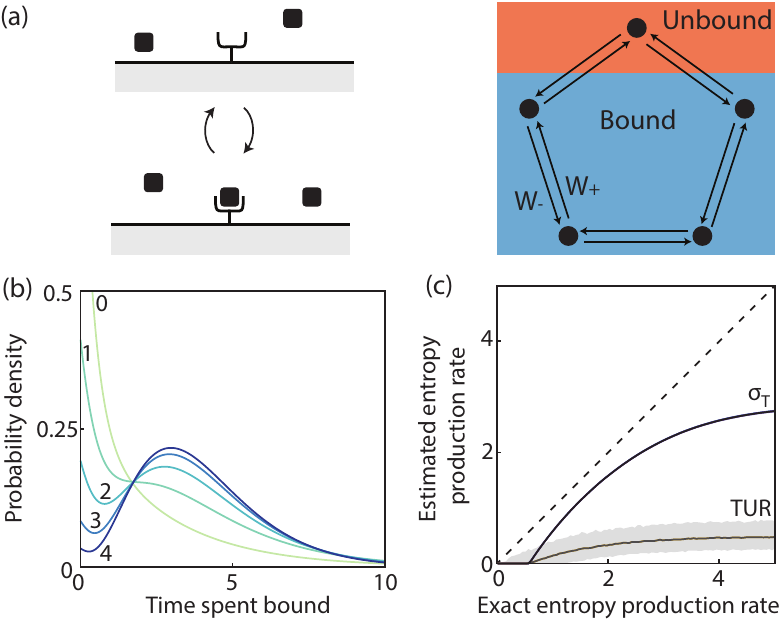}
\caption{\label{Fig:SensorFig}Bounding the EPR of a model active sensor.
(a) Cells sense through receptors on their surface, which detect concentration of some molecular
species (black squares) by switching between unbound (top left) and bound (bottom left) configurations. 
We consider a simple model of a sensor, with 4 internal states for the bound receptor configuration~\cite{harvey2020universal}. 
Clockwise transitions occur at a rate $W_{+}$, counter-clockwise with rate $W_{-}$.
(b) Distribution of time spent bound for $\sigma= 0,1,2,3,4$ with $k_B = 1 = W_{+}$,
$W_{-} \leq 1$.
(c) Estimates of the EPR as a function of the exact EPR, $\sigma$,
as calculated from 200 trajectories of length $T=2000$ for each value of $\sigma$. The $\sigma_T$  and TUR estimators
are shown with 95\% range of predictions.}
\end{figure}

To perform vital functions like chemotaxis, cells sense chemical 
concentration levels within their environment~\cite{Lan2012} through the stochastic binding and 
unbinding of molecules at surface receptors~\cite{harvey2020universal} [Fig.~\ref{Fig:SensorFig}(a)]. Recent works~\cite{Lang_2014,Lan2012,harvey2020universal} showed that active sensors can overcome the equilibrium Berg-Purcell sensing limit by expending energy, with
a trade-off between increased sensing accuracy, and energy expended. 
We apply Eq.~\eqref{e:sig_T} to a simple model of an active sensor~\cite{harvey2020universal}, with the
receptor system modeled as a ring with 5 states, one of which corresponds to
an unbound receptor, and with 4 internal states corresponding to a bound receptor [Fig.~\ref{Fig:SensorFig}(a)]. 
The rate of clockwise transitions is taken to be $W_{+}$, with counterclockwise transitions occuring 
at rate $W_{-}<W_{+}$, resulting in an EPR of $\sigma = \logfnInline{W_{+}}{W_{-}}$.
The $\sigma_T$ bound is reasonably close to the
exact value $\sigma$ in intermediate regimes of entropy production [Fig.~\ref{Fig:SensorFig}(c)]. For small values of 
$\sigma$, the same variance could be generated by an equilibrium process, whereas for large values
the variance could be generated at much lower cost by a system with more internal states~\cite{SM}. The original thermodynamic uncertainty relation (TUR)~\cite{TUR_proof_2016}, whilst a powerful tool for inference~\cite{Li_2019,Gingrich_natPhys}, cannot gain nontrivial bounds for this system, as there are no observed currents. A more recently introduced TUR~\cite{harvey2020universal,SM}, relating the mean and variance of the fraction of time spent in a metastate to the EPR, is the only existing estimator which we are aware of that can yield a nontrivial bound.
Figure~\ref{Fig:SensorFig}(c) shows that the TUR bound is substantially lower, 
and for a finite sample size has a much larger variance,  thus requiring a large amount of data for a reliable prediction compared 
to $\sigma_T$.

\par 
As the first of many experimental applications of $\sigma_T$, we consider stochastic gene regulatory networks~\cite{Jia2017,Bressloff2017}.
Cells regulate the intra-cellular concentration of proteins, changing the concentration in response to 
external stimuli, or maintaining a constant level by balancing production and 
degradation~\cite{Bressloff2017}. Proteins are created by enzymes translating mRNA, which in turn is transcribed
by enzymes from DNA instructions or genes~\cite{Paulsson2005}. Genes switch between active and inactive  metastates, the rate of switching regulated by protein concentration, amongst other things, and whilst active mRNA is transcribed~\cite{Jia2017} [Fig.~\ref{fig:TradeOff}(b)].
By examining the distribution of times which a gene spends inactive, we can deduce 
that the regulatory system is out-of-equilibrium. Recent experiments~\cite{Suter2011}  on  mammalian gene transcription measured the times for which certain genes
were active or inactive. From the histogram of inactivity periods [Fig.~\ref{fig:TradeOff}(b)]  for the Glutaminase gene, one finds $\langle t^2\rangle_{\text{off}}
/\langle t \rangle_{\text{off}}^2 = 1.6$ and $\tau = 0.9$h, so that  
$\sigma \geq 2.2 k_B/$h, whereas for the Bmal1 promotor, $\langle t^2 \rangle_{\text{off}} /
\langle t \rangle_{\text{off}}^2 = 1.5$ and $\tau = 1.1$h, resulting in $\sigma \geq 2.4 k_B/$h.
%%%%%%%%%%%%%%%% Fig 4
%%%%%%%%%%%%%%%% 
\par %\subsubsection{Behavioural states}
From complex neural networks to simple feedback loops, living systems make decisions on which behavior to execute,
and such decision making requires the expenditure of 
free energy~\cite{Mehta2012}. Recent experiments identified behavioral states of
different organisms~\cite{Cermak2020,Wan2018}. From dynamics on these metastates, it is possible to bound the EPR~\cite{Wan2018}. 
By attaching sensors to cows,  the authors of Ref.~\cite{TOLKAMP2010}  recorded whether the cows were standing
or lying, as well as the waiting time distribution of each [Fig.~\ref{fig:TradeOff}(c)]. 
From the time the cows spend lying (metastate~$A$), we can calculate a non-zero bound on the EPR. Specifically, for the three experiments involving
pregnant indoor-housed beef cows, out-wintered beef cows, and indoor-housed dairy cows~\cite{TOLKAMP2010}, we
find $\sigma_T = 4.1k_B/$h$, 3.2k_B/$h$, 2.4k_B/$h, respectively.
Assuming $k_B T \approx 9.9 \times 10^{-22}\,$Cal, we can therefore deduce that the cows consume at least 
$2.4 \times 10^{-21}\,$ Cal/h, in deciding whether to lie or stand. While this is a significant
underestimate, the $\sigma_T$-estimator does not assume any specific model of the decision making process,
and so does not require the cows to \lq possess spherical symmetry\rq~\cite{Stellman1296}.

%%%%%%%%%%%%%%%% Fig 4
\begin{figure*}
\includegraphics{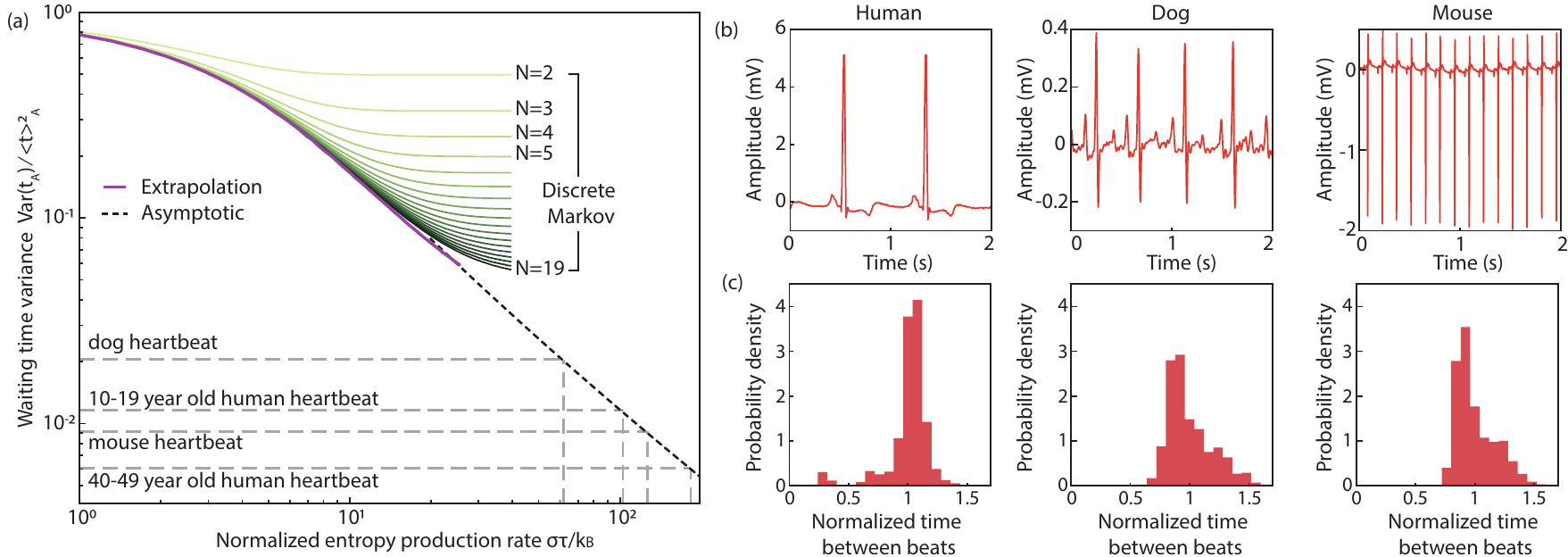}
\caption{\label{fig4}EPR in the small variance limit. (a) Numerical optimization (green) is only feasible for small $\sigma$, although extrapolating to $N = \infty$ extends this range (purple). Alternatively, the asymptotic limit of a continuous Langevin system extends beyond the range of  finite networks (dashed). Using the asymptotic relation, we can examine the entropic cost of
building an oscillator as precise as a heartbeat across different species (grey dotted). (b) Typical ECG measurements of heartbeats for humans, dogs and mice~\cite{PhysioBank,physioZoo}. (c) Distribution of time between beats, scaled so the mean time is 1, from a single experiment from each species. }
\end{figure*}
%%%%%%%%%%%%%%%% 

\par
The systems considered so far could be interpreted as timers; by paying an energetic cost they control the time spent in a subset of states
more precisely than an equilibrium system would allow. Previous studies of biological clocks assume measurements of time are performed by counting the number of full cycles around a circular network topology~\cite{Seifert_2016}, although such analysis can be extended to more general biological oscillators~\cite{Horowitz_clock,Junco2020}. 
However, the resulting TUR bound cannot be straightforwardly converted into bounds on $\text{Var}\, t_A$, and moreover the networks which saturate the TUR bound do no  better than equilibrium systems at minimizing $\text{Var}\, t_A$~\cite{SM}. To  elucidate the relationship between $\sigma$ and $\text{Var}\, t_A$, one can neglect the dynamics of metastate $B$ by letting $\langle t\rangle_B \to 0$, so that $\sigma \geq (2k_B/ \langle t \rangle_A )\Lambda ( \text{Var}\, t_A/\langle t \rangle_A^2) $. Our particular goal is to estimate a bound on the EPR of precise timers exhibiting small  variance $\text{Var}\, t_A/\langle t \rangle_A^2\ll 1$, such as heart beats.

\par
Unfortunately, it is not feasible to explore the limit $\text{Var}\,t_A \to 0$ numerically, as for a fixed number, $N$, of states in $A$, the largest precision possible is $\text{Var} \, t_A /  \langle t\rangle_A^2  \geq 1 / N$, and numerical minimization becomes prohibitively expensive for $N \gtrsim 20$~\cite{SM}. To gain analytical insight, we consider the continuum limit of an infinite number of internal states, $N\to \infty$, such that the dynamics in $A$ can be represented by a Langevin equation~\cite{SM,pavliotis2014stochastic}.  A continuous system can be approximated arbitrarily well by a discrete system~\cite{SM,Skinner2021,Chernyak_2006}, but a general discrete system need not be well approximated by a continuous Langevin equation~\cite{Busiello_2019,Herpich2020,Horowitz2015, SM}, so by searching over the space of Langevin dynamics we will find an upper bound on the true optimal precision \textit{vs.} entropy trade-off curve~\cite{SM}. In the limit of infinite precision, $\text{Var} \, t_A /  \langle t\rangle_A^2 \to 0$, we find analytically~\cite{SM} the asymptotic relation, $\text{Var} \, t_A /  \langle t\rangle_A^2 = 1/\hat{\sigma} + 4  \ln \hat{\sigma} / \hat{\sigma}^2 + o(\ln \hat{\sigma} / \hat{\sigma}^2)$, where $\hat{\sigma} = \sigma \langle t \rangle_A /2k_B$; achieving this in practice would require many internal states, a common trade-off in stochastic networks~\cite{Horowitz_clock,Paulsson2009Nature,PaulssonPRL2019}.  We apply this asymptotic prediction to heart beat data~\cite{Costa2005}  for humans~\cite{UMETANI1998} and other species \cite{physioZoo}.  Although the precision of the observed beating patterns renders numerical $\sigma_T$-estimates unfeasible,  the asymptotic formula implies entropic costs of at least  $280k_B/$s$, 480k_B/$s$, 270k_B/$s$, 2360k_B/$s to maintain the heartbeats for young humans, older humans, dogs, and mice, respectively (Fig.~\ref{fig4}).

\par%\subsubsection{Outlook and other examples}
In addition to the systems discussed above, non-equilibrium waiting time distributions have been measured for switching processes in cell fate decision making~\cite{Norman2013}, dwell times of kinesin motors~\cite{Isojima2016,Asbury2130}, cluster lifetimes of RNA polymerase~\cite{Cisse664}, the directional switching of the bacterial motor~\cite{BerryScience,Tu_2008}, swim-turn dynamics of algae and bacteria~\cite{THEVES2013,Polin2009}, direction reversal in swarming bacteria~\cite{Wu2009,Sliusarenko2006,Beer2013}, the repolarization times of migrating cancer cells~\cite{Zhou2020}, visual perception switching between metastable states~\cite{Krug2008}, and the duration of animal and insect flights~\cite{Elliott2009,raine2007}. The framework introduced here makes it possible to 
bound the extent to which these systems are out-of-equilibrium (Table S1) and, thereby, to quantify the trade-offs which biological systems are forced to make.

\begin{acknowledgments}
This work was supported by a MathWorks Fellowship (D.J.S.), a James S. McDonnell Foundation Complex 
Systems Scholar Award (J.D.), and the Robert E. Collins Distinguished Scholar Fund (J.D.). 
We thank the MIT SuperCloud and Lincoln Laboratory Supercomputing Center for providing HPC resources. The source code is available at \texttt{\href{https://github.com/Dom-Skinner/EntropyProductionFromWaitingTimes}{https://github.com/Dom-Skinner/ EntropyProductionFromWaitingTimes}}. 
\end{acknowledgments}

\end{document}

\raggedbottom
\title{Supplementary information:\\
Estimating entropy production from waiting time distributions}

\author{Dominic J. Skinner} 
\affiliation{Department of Mathematics, Massachusetts Institute of Technology, Cambridge Massachusetts 02139-4307, USA}
\author{J\"{o}rn Dunkel}
\affiliation{Department of Mathematics, Massachusetts Institute of Technology, Cambridge Massachusetts 02139-4307, USA}
\date{\today}
\maketitle
\tableofcontents

\section{Introduction and preliminaries}
We start with Markovian dynamics on a finite set of discrete states, \rev{$(1,\dots,N_T)$}, where a stochastic transition between states $i$ and $j$ occurs at a rate $W_{ij} \geq 0$. It follows that a probability distribution, $p_i(t)$ evolves through the master equation
\be
\frac{\d}{\d t} p_i = \sum_j p_j W_{ji}, 
\ee
with $W_{ii} = -\sum_{j\neq i} W_{ij}$. For irreducible, recurrent systems, meaning there exists a path of non-zero probability between any two states, the system has a unique \rev{steady-state distribution}, $\pi_i$, satisfying $\sum_j \pi_j W_{ji} = 0$ for all $i$, and all initial conditions will tend to this \rev{distribution}. The steady state average entropy production rate is\rev{
\be\label{eq:entP}
\sigma = k_B \sum_{i,j} \pi_i W_{ij} \log \frac{ \pi_i W_{ij}}{\pi_j W_{ji}}.
\ee}
Entropy production can be defined along single trajectories for systems starting with arbitrary probability distributions, but here we only consider this steady state expression~\cite{Gingrich_natPhys}. \rev{While we formally assume a mathematical non-equilibrium steady state (NESS), our bound on entropy production rate will extend to systems which are not a true NESS, but instead represent a quasi-stationary state (QSS)~\cite{Qian_2013,Skinner2021}. For an isothermal NESS at temperature $T$, $T\sigma$ can be interpreted as a physical heat, but in an isothermal QSS this interpretation no longer holds~\cite{Qian_2013}. However, $T\sigma$ retains the interpretation of a rate of free energy dissipation in both~\cite{Qian_2013}. Therefore, when there is a time scale separation between Markov dynamics on the discrete states and dynamics longer than the observation period, over which the transition rates may change, our estimator will be able to bound the entropy production rate, and $T \sigma_T$ will be a lower bound for the free energy dissipation rate~\cite{Qian_2013,Skinner2021}. }

\rev{An equivalent form of Eq.~\eqref{eq:entP}, is given by
\begin{align}\label{eq:entP2}
k_B \sum_{i,j} \pi_i W_{ij} \log \frac{ \pi_i W_{ij}}{\pi_j W_{ji}} &=
k_B \sum_{i,j} \pi_i W_{ij} \log \frac{ W_{ij}}{ W_{ji}} + k_B \sum_{i,j} \pi_i W_{ij} \log \frac{ \pi_i }{\pi_j } \\  \nonumber
&= k_B \sum_{i,j} \pi_i W_{ij} \log \frac{ W_{ij}}{ W_{ji}}  + k_B \sum_i \pi_i  \left[ \sum_j W_{ij} \right] \log \pi_i - k_B \sum_j \left[ \sum_i \pi_i W_{ij} \right] \log \pi_j \\ \nonumber
&=k_B \sum_{i,j} \pi_i W_{ij} \log \frac{ W_{ij}}{ W_{ji}} ,
\end{align}
since $\sum_j W_{ij} = \sum_i \pi_i W_{ij} = 0$, and this form is often stated as the entropy production rate~\cite{Gingrich_natPhys}.} We also note here that upon arrival at a state $i$, the distribution of time spent in that state is exponentially distributed with parameter $\lambda = - W_{ii}$.

Once the \rev{$N_T(N_T-1)/2$} transition rates $W_{ij}\geq 0$, $i\neq j$ are specified, the steady state distribution follows from the eigenvalue problem, and the entropy production rate can then be computed. It will later be useful to take an alternative formulation where we specify \rev{$N_T(N_T-1)/2$} rates of mass transfer, $n_{ij} = \pi_i W_{ij}$, together with the \rev{steady state distribution} $\pi_i$, and \rev{$N_T$} linear constraints $\sum_j n_{ij} = \sum_j n_{ji}$. After choosing $n_{ij}$ and $\pi_i$, we easily recover the transition rates as $W_{ij} = n_{ij}/\pi_i$, for $i \neq j$, $W_{ii} = - \sum_{j\neq i} W_{ij} = - \sum_{j\neq i} n_{ij}/\pi_i$. We see that the relation $\sum \pi_i W_{ij} = 0$ is satisfied, since this then becomes equivalent to the linear constraint on $n_{ij}$ which was imposed. From Eq.~\eqref{eq:entP2}, we see that now $\sigma$ depends only on the values of $n_{ij}$, and so $\pi_i$ can be specified independently. 

We now consider a system with two coarse grained \rev{metastates}. \rev{Typically, we will consider the case where only two coarse grained states are accessible because of  experimental constraints. However, we can also apply our framework if more than two metastates are accessible, as we are free to coarse grain further into just two states. Calling these metastates $A$  and $B$, we suppose} that the first \rev{$N$ micro-states} belong to the \rev{metastate} $A$, with \rev{$1\leq N < N_T$}. We can then write $W$ as a block matrix, such that
\be
W = \left( \begin{array}{cc} W_A & W_{AB} \\ W_{BA} & W_{B} \end{array} \right),
\ee
where $W_A$ is of size \rev{$N\times N$} and represents the within $A$ transition rates. We then have the following lemma from Ref.~\cite{rubino_sericola_1989}.
\begin{lemma}
For any such Markovian process in the stationary distribution, \rev{upon entering $A$, the distribution of time spent in $A$ before leaving} is given by $f_A(t) = \frac{1}{K} \pi_A W_A^2 \exp(W_A t) \bm{1}^\top$,
where $K = - \pi_A W_A \bm{1}^\top$ and $\bm{1}$ is a vector of ones.
\end{lemma}
From this lemma, it is straightforward to derive the moment statistics of the time spent in $A$, as 
\begin{align}
\langle t^k \rangle_A &= \int_{0}^\infty t^k f_A(t) \d t \\ \nonumber
&= \left[ t^k \frac{1}{K} \pi_A W_A^2 \exp(W_A t) \bm{1}^\top \right]_{t=0}^{\infty} - \int_0^\infty kt^{k-1} \frac{1}{K} \pi_A W_A \exp(W_A t) \bm{1}^\top \d t \\ \nonumber
%&= \cdots \\ \nonumber
&= (-1)^k k! \int_0^\infty \frac{1}{K} \pi_A W_A^{2-k} \exp(W_A t) \bm{1}^\top \d t \\ \nonumber
&= (-1)^{k+1} k! \pi_A W_A^{1-k} \bm{1}^\top/ K,
\end{align}
and similar for $B$.

\section{Non-monotonic waiting times imply out of equilibrium dynamics}
We rederive an expression for the waiting time distributions for a system in equilibrium, found earlier in Ref.~\cite{Tu_2008}. We start with the expression for the waiting time distribution for \rev{metastate} $A$, $f_A(t) = \frac{1}{K} \pi_A W_A^2 \exp(W_A t) \bm{1}^\top$, which holds regardless of whether detailed 
balance holds. If we further assume that the system is in
in thermal equilibrium so detailed balance applies, $\pi_i W_{ij} = \pi_j W_{ji}$, \rev{then $W$ is diagonalizable~\cite{van1992stochastic}. Specifically, we can define
a new matrix }
\be
D_{ij} = \frac{\sqrt{\pi_{Ai}} W^A_{ij}}{\sqrt{\pi_{Aj}}} = \frac{\sqrt{\pi_{Aj}} W^A_{ji}}{\sqrt{\pi_{Ai}}} = D_{ji},
\ee
which is symmetric by construction. Therefore, $D$ admits a spectral decomposition, 
\be
D_{ij} = \sum_k \lambda_k w^k_i w^k_j,
\ee
where $\lambda_k$ are the real eigenvalues, and $w^k$ are the orthonormal eigenvectors. Since $D$ is related to $W$
through a similarity transformation, $D= \Lambda W \Lambda^{-1}$, with $\Lambda_{ij} = \delta_{ij} \sqrt{\pi_{Ai}}$, 
we have that 
\be
f_A(t) = \frac{1}{K} \pi_A \Lambda^{-1} D^2 \exp(Dt) \Lambda \bm{1}^\top,
\ee
where $\left[ D^2 \exp(Dt) \right]_{ij} = \sum_k \lambda_k^2 e^{\lambda_k t} w_i^k w_j^k$. Therefore,
\begin{align} \nonumber
f_A(t) &= \frac{1}{K} \sum_{i,j} \; \sqrt{\pi_{Ai}} \left( \sum_k \lambda_k^2 e^{\lambda_k t} 
w_i^k w_j^k \right) \sqrt{\pi_{Aj}} \\
 &= \frac{1}{K} \sum_k \lambda_k^2 e^{\lambda_k t} \langle \sqrt{\pi_A}, w^k \rangle ^2.
\end{align}
As this represents a waiting time probability distribution which is normalizable, $f_A \to 0$ as $t\to \infty$, and so $\lambda_i <0$, but
$\lambda_k^2 \langle \sqrt{\pi_A}, w^k \rangle^2 \geq 0$, so $f_A(t)$ is the sum of decaying exponentials, each
with a non-negative weight. There are several natural corollaries of this result. As noted by Ref.~\cite{Tu_2008}, 
this implies that waiting time distributions of equilibrium systems are monotonically decaying functions, and all derivatives
are monotonic as well (see Eq. (1) and (2) in Ref~\cite{Tu_2008}). Here, we also note that the requirement that $f_A = \sum_i p_i |\lambda_i| \exp(-|\lambda_i| t)$,
for $p_i >0$, $\lambda_i <0$, $\sum_i p_i = 1$, implies the moments $\langle t^n \rangle_A = n! \sum_i p_i/|\lambda_i|^n$, and
hence the Cauchy-Schwarz inequality implies that 
\be
\left(\sum_i p_i/|\lambda_i| \right)^2 \leq \left(\sum_i p_i/\lambda_i^2 \right) \left(\sum_i p_i \right),
\ee
and so $2 \langle t \rangle_A^2 \leq \langle t^2 \rangle_A$ for equilibrium systems.

\section{Numerical minimization}
As discussed in the main text, we focus on the numerical minimization problem of finding the curve
\be
\Lambda(\theta) = \min_{\mathcal{R}} \{\sigma(\mathcal{R})/k_B | \langle t \rangle_{A,\mathcal{R}} = \langle t \rangle_{B,\mathcal{R}} = 1, \langle t^2 \rangle_{A,\mathcal{R}} = \theta \},
\ee
which yields the bound 
\be
\sigma \geq \frac{2k_B}{\langle t \rangle_A + \langle t \rangle_B} \Lambda\left( \frac{\langle t^2 \rangle_A}{\langle t \rangle_A^2} \right),
\ee
for an arbitrary system. More specifically, finding $\Lambda$ results in the following constrained minimization problem,
\begin{align}
\Lambda(\theta) &= \min_{n,\pi} \sum_{i < j} \logfn{n_{ij}}{n_{ji}}, \\
\nonumber & s.t. \ \pi_{A} \bm{1}^\top = \pi_B \bm{1}^\top = - \bm{1} n_{A} \bm{1}^\top = 1/2, \\
\nonumber & -4\pi_A n_A^{-1} \pi_A^\top = \theta, \\
\nonumber & \pi \geq 0,\  n_{ij} \geq 0,\  n_{ii} = - \sum_{j\neq i } n_{ij} = -\sum_{j \neq i} n_{ji}, 
\end{align}
where the minimum is taken over all possible network topologies.
We consider the subproblem \rev{$\Lambda_{N,M}(\theta)$}, where the minimum is taken over \rev{$N$} internal states of $A$ 
(we take these as \rev{$i=1,\dots, N$}), and $M$ internal states of $B$. We make the following claim
\begin{lemma}
The minimum is achieved when only 1 internal state of B, ($M=1$), is used.
\end{lemma}
To see this, start from any solution that satisfies the constraints with $M$ internal states of $B$. 
Then take \rev{$(\pi_{N+1},\dots,\pi_{N+M}) \mapsto \hat{\pi}_{N+1} = 1/2$},
and for \rev{$i\leq N$}, \rev{$(n_{i,N+1},\dots, n_{i,N+M}) \mapsto \hat{n}_{i,N+1} = \sum_j n_{i,N+j}$},
similar for \rev{$\hat{n}_{N+1,i}$},
and \rev{$\hat{n}_{N+1,N+1} = - \sum_{i \leq N} \hat{n}_{i,N+1}$}. This new system satisfies the constraints, 
as the $\theta$ constraint does not depend on $n_B$. To see that the entropy production rate does not 
increase, \rev{note that
\begin{align}
\sum_{i < j} \logfn{n_{ij}}{n_{ji}}  
 &\geq  \left( \sum_{i < j \leq N} + \sum_{i \leq N, j > N} \right) \logfn{n_{ij}}{n_{ji}}\\
&\geq \nonumber \sum_{i < j \leq N} \logfn{n_{ij}}{n_{ji}} + \sum_{i\leq N} \logfn{\sum_{j>N} n_{ij} }{\sum_{j>N} n_{ji}} \\
\nonumber & = \sum_{i < j \leq N} \logfn{\hat{n}_{ij}}{\hat{n}_{ji}} + \sum_{i \leq N}\logfn{ \hat{n}_{i,N+1} }{\hat{n}_{N+1,i}}, 
\end{align}}
\noindent where we have used the convexity of the function $f(x,y) = \logfnInline{x}{y}$.
Therefore, the new system $\hat{n}$ produces entropy at a rate less than or equal to the original system, and
so it is unneccessary to minimize over multiple internal states of $B$.
Therefore, from now on, we compute numerically \rev{$\Lambda_N(\theta)$}, where the minimum is taken over \rev{$N$} internal
states of $A$ and a single internal state of $B$.

To perform numerical minimization, we perform a global optimization using \texttt{MATLAB}'s \texttt{fmincon} 
function~\cite{MATLAB:2019}, together with a global search strategy~\cite{GlobalSearch}, with
random initial conditions. Since the curves $\Lambda_N(\theta)$ are monotonic, finding it is equivalent to 
fixing $\sigma$ and minimizing $\theta$, finding instead a curve $\theta = \Gamma_N(\sigma)$. We chose to solve this problem instead, as the minimization was found to 
convergence more consistently, although both approaches are ultimately equivalent. We sample feasible initial 
conditions by first sampling a random matrix with positive entries, and successively projecting onto constraint
subspaces until convergence, where projection onto the entropy constraint is done approximately with a Newton step.
We also supply the minimization algorithm with the gradient and Hessian information of the function and 
constraints, which we briefly derive here.
We take $n^A$, $\pi_A$ as our variables, with the remaining variables determined by the constraints, where $n^A_{ij} = n_{ij} = \pi_i W_{ij}$ for $i,j \leq N$. We therefore
must consider the gradient and Hessian of
\begin{subequations}
\begin{align}
\sigma(n^A) &= \sum_{i<j} \logfn{n^A_{ij}}{n^A_{ji}} \ + \sum_{i} \logfn{c_i}{d_i}, \\
f(n^A,\pi_A) &= \pi_A (n^A)^{-1} \pi_A^\top,
\end{align}
\end{subequations}
with $c_i = -\sum_j n^A_{ij}$, and $d_i = -\sum_j n^A_{ji}$, and we have defined $f  = -\theta/4$ 
for clarity. We first note that
\be
\frac{\partial [ (n^A)^{-1}]_{ij}}{\partial n^A_{pq}} = - [(n^A)^{-1}]_{ip} [(n^A)^{-1}]_{qj},
\ee
and define $y = (n^{A})^{-1} \pi_A$, $z = (n^{A})^{-T} \pi_A$, so that the first order derivatives in $f$ are
\begin{subequations}
\begin{align}
\frac{\partial f}{\partial \pi_{Ai}} &= \left( [(n^A)^{-1}]_{ij} + [(n^A)^{-1}]_{ji} \right) \pi_j = y_i + z_i, \\
\frac{\partial f}{\partial n^A_{ij}} &= - \pi_{Ai} [(n^A)^{-1}]_{ip} [(n^A)^{-1}]_{qj} \pi_{Aj} = - z_p y_q,
\end{align}
\end{subequations}
and second order derivatives are
\begin{subequations}
\begin{align}
\frac{\partial^2 f}{\partial \pi_{Ai} \partial n^A_{pq}} &= - \left[
[(n^A)^{-1}]_{ip} [(n^A)^{-1}]_{qj}  + [(n^A)^{-1}]_{jp} [(n^A)^{-1}]_{qi} \right] \pi_{Aj} \\ \nonumber
&= -[(n^A)^{-1}]_{ip} y_q - z_p [(n^A)^{-1}]_{qi} \\ 
\frac{\partial^2 f}{\partial \pi_{Ai} \partial \pi_{Aj}} &= [(n^A)^{-1}]_{ij} + [(n^A)^{-1}]_{ji} \\ \nonumber
\frac{\partial^2 f}{\partial n^A_{pq} \partial n^A_{rs}} &=
\pi_{Ai} \left[ [(n^A)^{-1}]_{ir} [(n^A)^{-1}]_{sp} [(n^A)^{-1}]_{qj}  + [(n^A)^{-1}]_{ip} [(n^A)^{-1}]_{qr} [(n^A)^{-1}]_{sj} \right] \pi_{Aj} \\
&= z_r [(n^A)^{-1}]_{sp} y_q + z_p [(n^A)^{-1}]_{qr} y_s.
\end{align} 
\end{subequations}
Moving to $\sigma$, for the purposes of numerical stability, we replace $\log(a/b)$ with $\log( (a+\epsilon)/(b+\epsilon))$, where $\epsilon$ is a small numerical parameter
which ensures that the constraint and derivatives remain finite at the boundaries. We introduce the functions
\be
g(a,b) = g(b,a) = (a-b)\log\left( \frac{a + \epsilon}{b+\epsilon} \right),
\ee
and
\be
h(a,b) = \frac{\partial g}{\partial a} = \log\left( \frac{a + \epsilon}{b+\epsilon} \right) +
\frac{a - b}{a + \epsilon},
\ee
so that,
\be
\sigma = \sum_{i < j} g(n^A_{ij},n^B_{ji}) \ + \sum_{i} g\left( c_i, d_i \right),
\ee
where we recall that $c_i = -\sum_j n^A_{ij}$, and $d_i = -\sum_j n^A_{ji}$.
To calculate the derivatives, we make use of
\be
\frac{\partial c_i}{\partial n^A_{jk}} = -\delta_{ij}, \qquad
\frac{\partial d_i}{\partial n^A_{jk}} = -\delta_{ik},
\ee
to find
\be
\frac{\partial \sigma }{\partial n^A_{ij}} =  h(n^A_{ij},n^A_{ji}) - h(c_i,d_i) - h(d_j ,c_j),
\ee
Defining the second derivative functions
\begin{subequations}
\begin{align}
h_a &= \frac{\partial h}{\partial a} = \frac{a+b+2\epsilon}{(a+\epsilon)^2}, \\
h_b &= \frac{\partial h}{\partial b} = -\frac{a+b+2\epsilon}{(a+\epsilon)(b+\epsilon)},
\end{align}
\end{subequations}
we have that
\begin{align} \nonumber
\frac{\partial^2 \sigma }{\partial n^A_{ij} \partial n^A_{pq}} =& 
h_a(n^A_{ij},n^A_{ji}) \delta_{ip}\delta_{jq} 
+ h_b(n^A_{ij},n^A_{ji}) \delta_{iq}\delta_{jp} \\ 
&+ \delta_{ip}h_a(c_i,d_i) + \delta_{iq}h_b(c_i,d_i)  \delta_{jq} h_a(d_j,c_j) + \delta_{jp}h_b(d_j,c_j)
\end{align}
\section{Asymptotic solution}
Here, we derive analytically the small and large entropy asymptotics for a system with a fixed number, \rev{$N$}, internal states within $A$. We use the asymptotic expansion notation $f(x) \sim a_0 f_0(x) + a_1 f_1(x)$ to mean that $f - a_0 f_0  = o(f_0)$, and $f - a_0 f_0 - a_1 f_1 = o(f_1)$ in the limit.
\subsection{Small entropy asymptotics}
\begin{figure}
\includegraphics{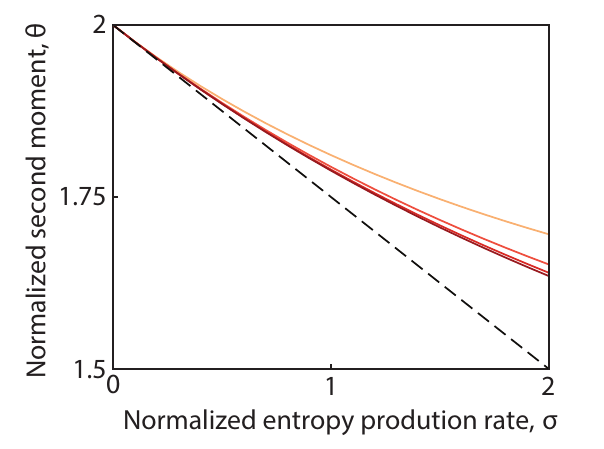}
\caption{\label{fig:small}Small entropy asymptotics, for $\sigma \ll 1$, predicts $\theta \sim 2 - \sigma/4$ (black 
dotted line), for the cannonically
rescaled system. This agrees well with numerical results shown in shades of red for $N= 2,3,4,5$ internal
states, which rapidly converge to a single curve.}
\end{figure}

Numerically we find that the small entropy asymptotic bound is achieved with \rev{$N=2$}. Here, we find the asymptotic
form $\theta \sim 1 - \sigma/16$ for $\sigma \ll 1$. We take 
\be
n_A = \left[ \begin{array}{cc} A_{11} & A_{12} \\ A_{21} & A_{22} \end{array} \right],
\ee
so that we must minimize
\be
\theta = \frac{-4}{A_{11}A_{22} - A_{12} A_{21}} \pi_A^\top 
\left[ \begin{array}{cc} A_{22} & -A_{12} \\ -A_{21} & A_{11} \end{array} \right] \pi_A,
\ee
subject to 
\begin{align}
\sigma =& \logfn{A_{12}}{A_{21}} - (A_{12} - A_{21}) \log \left( \frac{A_{11} + A_{12}}{A_{11} + A_{21}} \right) - (A_{12} - A_{21}) \log \left( \frac{A_{22} + A_{12}}{A_{22} + A_{21}} \right) \\\nonumber
-\frac{1}{2} =& A_{11} + A_{12} + A_{21} + A_{22},
\end{align}
in addition to the inequality constraints
\begin{subequations} 
\begin{align}
A_{11},A_{22} &\leq 0, \\ 
A_{12},A_{21} &\geq 0, \\
A_{11} + A_{12}, \ A_{11} + A_{21} \ A_{22} + A_{21}, \ A_{22} + A_{12} &\leq 0.
\end{align}
\end{subequations}
Let
\begin{align}
A_{12} =  \alpha + \beta , \qquad A_{21} =  \alpha - \beta,
\end{align}
and we assume that $\alpha = O(1)$. 
We have that
\be
\theta = \frac{-4}{A_{11}A_{22} -\alpha^2 + \beta^2} \pi_A^\top 
\left[ \begin{array}{cc} A_{22} & -\alpha \\ -\alpha & A_{11} \end{array} \right] \pi_A,
\ee
where, since $\pi_{A1} + \pi_{A2} = 1/2$, we have that
\begin{align} \nonumber
\theta &= \frac{ - A_{22} + 4(A_{22} + \alpha) \pi_{A2} - 4(A_{11} + A_{22} + 2\alpha) \pi_{A2}^2}{A_{11}A_{22} -\alpha^2 + \beta^2},\\
 &= \frac{2\pi_{A2}^2 +4(A_{22}+\alpha)\pi_{A2} - A_{22}}{A_{11}A_{22} -\alpha^2 + \beta^2},
\end{align}
since $A_{11} + A_{22} + 2 \alpha = -1/2$.
This minimized when $\pi_{A2} = -(A_{22}+\alpha)$, so that
\be
\theta = \frac{2A_{11} A_{22} - 2\alpha^2}{A_{11} A_{22} - \alpha^2 + \beta^2} \sim
2 - \frac{2\beta^2}{A_{11}A_{22} - \alpha^2}.
\ee
The leading order expression for the entropy production rate is,
\be
\sigma = \frac{4\beta^2}{\alpha} - \frac{4\beta^2}{A_{11} +\alpha} - \frac{4\beta^2}{A_{22} + \alpha}.
\ee
Since we can interchange $A_{11}$, $A_{22}$ without changing the function or constraints, the minimum has
$A_{11} = A_{22}$, so that $A_{11} = -\alpha - 1/4$, and
\be
\sigma = 4\beta^2 ( 1/\alpha +8),
\ee
so
\be
\theta \sim 2 - \frac{8\alpha\sigma}{(1 + 8\alpha)^2},
\ee
which is minimized when $\alpha = 1/8$, giving 
\be \theta \sim 2 - \sigma/4. \ee
As a consistency check, we observe
\begin{align}
n_A = \left[ \begin{array}{cc} -\frac{3}{8} &  \frac{1}{8} + \frac{\sqrt{\delta}}{8} \\
 \frac{1}{8} - \frac{\sqrt{\delta}}{8} & -\frac{3}{8} \end{array} \right]  , \qquad
c  = \left[ \frac{1}{4} - \frac{\sqrt{\delta}}{8} ,\frac{1}{4} + \frac{\sqrt{\delta}}{8}  \right] , \qquad
d  = \left[ \frac{1}{4} + \frac{\sqrt{\delta}}{8} ,\frac{1}{4} - \frac{\sqrt{\delta}}{8}  \right]
\end{align}
which produces entropy at a rate
\begin{align} \nonumber
\sigma &= \frac{\sqrt{\delta}}{4} \log \left( \frac{ 1 + \sqrt{\delta}}{1 - \sqrt{\delta}}\right) +
 \frac{\sqrt{\delta}}{2} \log \left( \frac{ 1 + \sqrt{\delta}/2}{1 - \sqrt{\delta}/2}\right) = \delta + O(\delta^3),
\end{align}
while
\be
\theta = \frac{2}{1+\delta^2/8} \sim 2 - \delta^2 /4 + O(\delta^4)
\ee
This asymptotic solution agrees well with numerics for $\sigma \ll 1$, Fig.~\ref{fig:small}.

\subsection{Large entropy asymptotics}
In the limit of infinite entropy production \rev{with a fixed number, $N$, of internal states in $A$}, we can choose a deterministic path through the internal states,
but uncertainty in the waiting time distribution still occurs due to the exponentially distributed dwell times
in each internal state. However, the total uncertainty for the total wait duration can be minimized by choosing
a path that sequentially moves through each internal state, so that 
\be
n^A = \frac{1}{2} \begin{bmatrix} 
    -1 & 1 &  \\
     & -1 &  1  \\
    & & \ddots & \\
     & &       & -1 & 1 \\ 
     & &       & & -1 
    \end{bmatrix},
\ee
with $c = (0,\dots ,0,1/2)$, $d = (1/2,0,\dots,0)$, and 
\be
(n^A)^{-1} = -2\begin{bmatrix} 
    1 & 1 & &\cdots & 1  \\
     & 1 &  1  \\
    & & \ddots &  & \vdots\\
     & &       & 1 & 1 \\ 
     & &       & & 1 
    \end{bmatrix},
\ee
and \rev{$\pi_A = \bm{1}/(2N)$}, so \rev{$\theta = 1 + 1/N$}. Thus with more states we can 
reach $\theta= 1$ in the \rev{$N\to \infty$} limit, corresponding to zero variance in wait time.
Taking a perturbation from this solution, we find that to leading order
\begin{align}
\delta \theta =&\ 4\pi_A^\top (n^A)^{-1} \delta n^A (n^A)^{-1} \pi_A - 4\pi_A^\top (n^A)^{-1} \delta \pi_A -4 \delta \pi_A^{T} (n^A)^{-1} \pi_A \\ \nonumber
		 =&\ 4\pi_A^\top (n^A)^{-1} \delta n^A (n^A)^{-1} \pi_A + \pi_A^\top ((n^A)^{-1} + (n^A)^{-T}) \delta \pi_A \\ \nonumber
         =&\ 4\pi_A^\top (n^A)^{-1} \delta n^A (n^A)^{-1} \pi_A,
\end{align}
since \rev{$\pi_A = \bm{1}/(2N)$}, and $[(n^A)^{-1} +(n^A)^{-T}]_{ij} = -2 - 2\delta_{ij}$, so the 
condition $\bm{1}^\top \delta \pi_A =0$ causes
the term linear in $\delta \pi_A$ to vanish. We can write 
\be
\delta \theta = 4 \sum_{i,j} X_{ij} \delta n^A_{ij},
\ee
where \rev{
\be
X_{ij} = \frac{1}{N^2} i (N+1 -j).
\ee}
That this linear term does not vanish is due to it being on the boundary of the feasible set. 
The entropy production rate to leading order is \rev{
\begin{align}\nonumber
\sigma &= \sum_{i=1}^{N-1} \ -\frac{1}{2}\log \left( 2\delta n^A_{i+1,i}\right)
 - \frac{1}{2} \log \left[ -2\sum_j \delta n^A_{1j}  \right]  - \frac{1}{2} \log \left[ -2\sum_j \delta n^A_{jN} \right].
\end{align}}
We solve the perturbation problem by fixing $\delta \theta = \epsilon$, and minimize $\sigma$, with
$\delta n^A_{ij} = \epsilon Y_{ij}$, so that we solve \rev{
\begin{subequations}
\begin{align} 
\min_Y \sigma(Y) &= \frac{N+1}{2}\log (1/2\epsilon) - \frac{1}{2}\sum_{i=1}^{N-1} \log (Y_{i+1,i}) 
- \frac{1}{2}\log\left[ -\sum_j Y_{1j} \right]
- \frac{1}{2}\log \left[ -\sum_j Y_{jN} \right]\\
 \text{s.t.} &\ 4\sum_{i,j} X_{ij} Y_{ij} = 1, \\
& \sum_{i,j} Y_{ij} = 0 \\
& \sum_{j} Y_{ij} \leq 0, \text{ for } i = 1,\dots, N-1 \\
& \sum_{j} Y_{ji} \leq 0 , \text{ for } i = 2,\dots, N\\
& Y_{ij} \geq 0, \text{ for } i \neq j \text{ and } i \neq j-1  
\end{align}
\end{subequations}}

\begin{figure}
\includegraphics[width=0.9\textwidth]{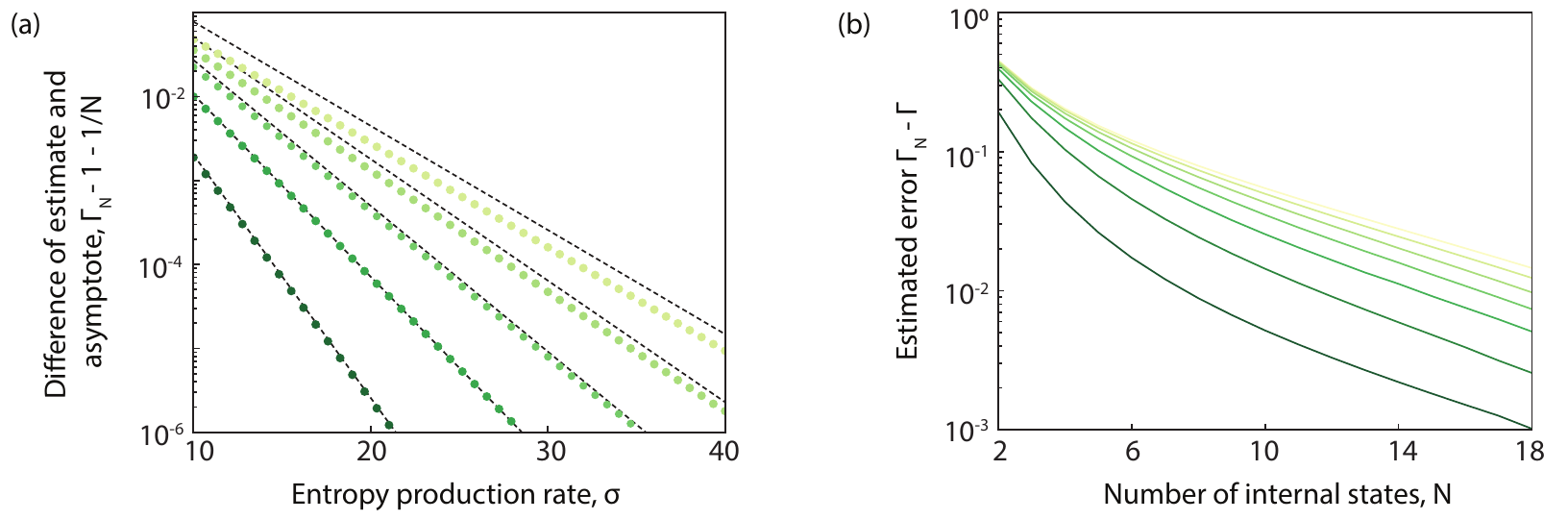}
\caption{\label{fig:large_n}
(a) Numerical convergence of $\Gamma_N - 1 - 1/N \to 0$ as $\sigma \to \infty$, against predicted asymptotic formula, Eq.~\eqref{eq:large_as}. Curves for $2\leq N \leq 6$ are shown from bottom to top (green points) together with respective asymptotic predictions (dotted lines). As $N$ increases, larger values of $\sigma$ are required for the asymptotic regime to hold. 
(b) Exponential convergence of $\Gamma_N$ to our estimate of $\Gamma$, for $\sigma = 5,10,
\dots,35$ bottom to top. Values of $\Gamma_N$ from $N=12,\dots, 18$ were used to fit the estimate of $\Gamma$.}
\end{figure}

We can already predict that \rev{$\sigma = \frac{1}{2}(N+1) \log (1/2\epsilon) + C$}, where $C = O(1)$, 
to find the asymptotic formula, \rev{
\be
\theta(\sigma) \sim 1 + \frac{1}{N} + \frac{1}{2}\exp\left( \frac{C - \sigma}{N+1} \right)
\ee}
in the large $\sigma$ limit. To find $C$, we take $Y_{i,i} \neq 0$, $Y_{i+1,i} \neq 0$, and $Y_{ij} =0$ 
otherwise, and suppose the constraints on row
and column sums on $Y_{ij}$ are tight. Then 
\be
Y = \begin{bmatrix} 
    -b &    &   &         &    & \\
    a     & -a  &   &         &    & \\
           & a & -a &         &    & \\
           &    &   &  \ddots &    & \\
           &    &   &         & -a  &  \\
           &    &   &         & a & -c\\
    \end{bmatrix},
\ee
where $a,b,c>0$ are some variables to be minimized over.
Since $b$ and $c$ can be interchanged without changing the objective or constraints,
and the current linearized problem is convex, we may take $b=c$, so that $b=c=a/2$
from the constraint $\sum_{i,j} Y_{ij} = 0$.
The constraint $4\sum_{i,j} Y_{ij} X_{ij} = 1$ gives \rev{
\begin{align} 
1 &= \frac{4}{N^2}\left[-a\sum_{i=1}^{N-1}i(N+1-i) + a\sum_{i=1}^{N-1}(i+1)(N+1-i) \right]  =  \frac{2a (N^2+N+2)}{N^2}
\end{align}}
so that \rev{
\be
a = \frac{N^2}{2(N^2 + N -2)},
\ee}
and hence \rev{
\begin{align}
\sigma =& \frac{N+1}{2} \log (1/2\epsilon)  - \frac{N+1}{2} \log \left( \frac{N^2}{2(N^2 + N -2)} \right) 
+ \log 2
\end{align}}
meaning that \rev{
\be\label{eq:large_as}
\theta(\sigma) \sim 1 + \frac{1}{N} + \left( \frac{ N^2 + N -2}{N^2} \right) 2^{2/(N+1)} 
\exp\left( -\frac{2\sigma}{N+1} \right).
\ee}
This agrees well with with numerical results, Fig.~\ref{fig:large_n}(a), although for larger \rev{$N$}, 
the asymptotic correction to \rev{$\theta = 1 + 1/N$} only holds for very large $\sigma$. This means that for larger \rev{$N$}, this asymptotic formula
does not provide much insight into the intermediate values of $\sigma$, and so can not replace performing a full numerical minimization within those regions.

\section{Numerical convergence}
To determine the exact bound $\Lambda(\theta)$, we need to take the limit, $N\to \infty$ of infinitely many
internal states. Of course, this is not practical numerically, so we must verify that finite $N$ values
converge, and estimate their error. Recall that we find sucessive values of $\theta$ for fixed $\sigma$,
calling these $\Gamma_N(\sigma) = \Lambda_N^{-1}(\sigma)$.
We see convergence appears exponential, 
\be
\Gamma_N \approx \Gamma + \Gamma_0 e^{-k N},
\ee
from which we wish to estimate $\Gamma$ from finite $N$ data. To do so, we define a residual
\be
R = \sum_{N=M}^{N=L} \left( \log [ \Gamma_N - C_1] - C_2 - C_3 N \right)^2,
\ee 
where by minimizing the residual with respect to $C_1,C_2,C_3$, we find the best exponential fit,
and can estimate $\Gamma \approx C_1$. Calling $y_N = \log[ \Gamma_N - C_1 ]$, and $x_N = N$, finding
the values of $C_2$, and $C_3$ that minimize the residual is straightforward linear regression,
\be
\begin{aligned}
C_3 &= \frac{\sum_N (x_N - \bar{x}) ( y_N - \bar{y} )}{\sum_N (x_N - \bar{x})^2}, \\
C_2 &= \bar{y} - C_3 \bar{x},
\end{aligned} 
\ee
The resulting residual is a non-linear function of $C_1$, and so we find the minimum numerically.
This serves as an estimate for $\Gamma$, and indeed we see exponential convergence to zero of $\Gamma_N - \Gamma$,
Fig.~\ref{fig:large_n}(b). Moreover, the discrepancy between $\Gamma_N$ and our estimate for $\Gamma$ serves as an 
order of magnitude estimate for the error of our $\Gamma$ estimate. This ranges from $10^{-3}$ for $\sigma = 5$,
to $1.5\times 10^{-2}$ for $\sigma = 35$, which is small compared to the other sources of error introduced here by
estimates of the mean and second moment of the waiting time distribution.
\section{Thermodynamic Uncertainty Relation}
Given two states $i$ and $j$, suppose $J_T(i,j)$, counts the net number of transitions $i\to j$ in some time $T$, then we can define a coarse grained current, $J_T = \sum_{i<j} d(i,j) J_T (i,j)$ for some weights $d$, with $d(i,j) = - d(j,i)$. The thermodynamic uncertainty relation (TUR) states that $\sigma T \geq 2k_B\langle J_T \rangle^2/\text{Var}\, J_T$, for any time $T$~\cite{Gingrich_natPhys}. Whilst this has proved a valuable tool for inference~\cite{Gingrich_natPhys,Seifert_AnnRev}, for our two macro-state systems, no coarse grained currents ($\langle J _T \rangle \neq 0$) can be observed. Recently however, an extension of this inequality has been derived~\cite{harvey2020universal}, which states that for two state systems, \rev{
\be\label{eq:TUR}
\sigma \geq \frac{8 \langle q_A \rangle^2 \langle q_B \rangle^2}{T \text{Var}\, q_A} - \frac{4}{\langle t \rangle_A + \langle t \rangle_B},
\ee
where $q_A$ is the empirically observed fraction of time spent in $A$ for an observation of length $T$, and $\langle q_A \rangle = r_A$, $\text{Var}\, q_A$ are the average and variance of this quantity respectively, similar for $q_B$. In Ref.~\cite{harvey2020universal}, the result is stated in terms of the quantity 
\begin{equation}
R^\pi = \sum_{i\in A, j \in B} n_{ij} = \sum_{i\in A, j \in B} n_{ji},
\end{equation}
but since 
\be
\langle t \rangle_A = \frac{\pi_A \bm{1}^\top}{ - \pi_A W_A \bm{1}^\top}  = \frac{\pi_A \bm{1}^\top}{ R^\pi},
\ee
this can be rewritten as $R^\pi = 1/(\langle t \rangle_A + \langle t \rangle_B)$, which we have done in Eq.~\eqref{eq:TUR}.} This TUR does not hold for finite $T$, but holds in the limit $T \to \infty$. 

\subsubsection{Active sensor}
To compare against this result for the active sensor example, we simulated 200 long trajectories of length $T=2000$ so the trajectory completes $O(400)$ cycles, and computed this modified TUR bound by calculating the average fraction of time spent in $A$ and $B$, as well as the variance of these across trajectories, as described in the main text. \rev{We find that the $\sigma_T$ bound is significantly improved on the TUR bound, both in terms of the value of the bound as well as the uncertainty, Fig. 3. It is worth noting that as $\sigma \to \infty$ in the active sensor, $\langle t^2 \rangle_A/\langle t \rangle_A^2 \to 1 + 1/4$, since in our model there are 4 internal states of $A$, yet such a value of $\langle t^2 \rangle_A/\langle t \rangle_A^2$ could be achieved at a far lower $\sigma$ by utilizing a system with more internal states, which is why $\sigma_T$ results in a significant underestimate for large $\sigma$. For small enough $\sigma$, the statistics satisfy $\langle t^2 \rangle_A/\langle t \rangle_A^2 \geq 2$, which could have conceivably been generated by an equilibrium process, again resulting in a significant underestimate by $\sigma_T$. However, in the intermediate regime, which the system would operate in, the bound is reasonably close.}

\rev{
\subsubsection{Random transition rates}
To further compare the TUR estimator with $\sigma_T$, we generated 200 systems each containing three states and with the transition rates $W_{ij}$ of each system drawn from an exponential distribution with mean 1. We then coarse-grained, taking the first two states as part of the metastate $A$ and the remaining state as part of $B$. For each of these systems, we computed both the TUR bound and the $\sigma_T$ bound from 50 trajectories of length $T=2000$. Plotting the two bounds against each other reveals that the $\sigma_T$ bound is typically a significant improvement on the (median) TUR bound, Fig.~\ref{fig:TURComp}. Moreover, when repeatedly calculating these bounds from simulations, the 95\% range of predictions is small for $\sigma_T$, whereas for most of the TUR bounds, the 95\% range contains zero, and so can not reliably identify the system as producing entropy at all, Fig.~\ref{fig:TURComp}.
\begin{figure}
\includegraphics{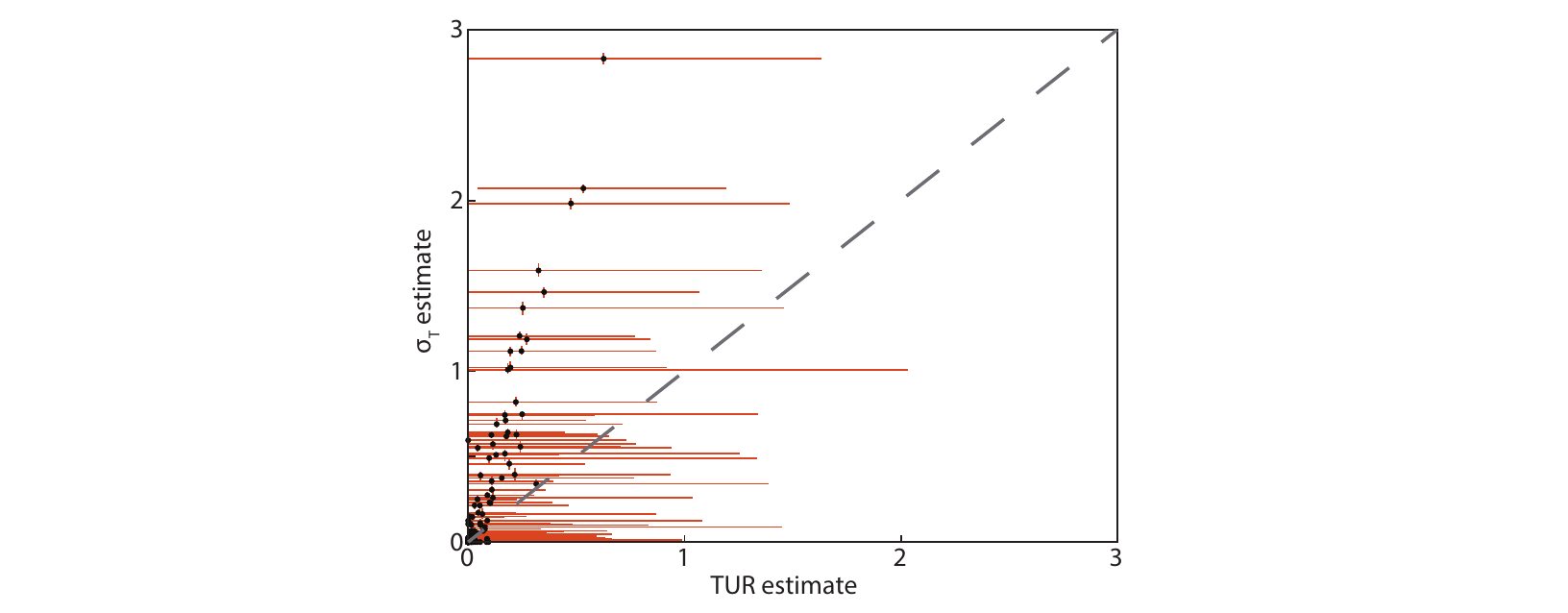}
\caption{\label{fig:TURComp}\rev{Entropy production bounds for random systems for both TUR and $\sigma_T$. The value of the TUR bound with $k_B=1$ is plotted against the value of the $\sigma_T$ bound. For each randomly generated transition matrix, the bounds were computed from 50 trajectories of length $T=2000$ (black points indicate sample medians). Error bars in red show 95\% range of predictions for both TUR and $\sigma_T$, and were calculated by repeatedly calculating the bounds 400 times.}}
\end{figure}
}

\section{Continuous Langevin system}
Gaining some understanding of the behavior of the curve $\theta = \Gamma(\sigma)$ in the large $\sigma$ limit is important to understanding the precision limits of stochastic clocks and timers. When time is measured by counting the number of full rotations around some circular topology, the thermodynamic uncertainty relation (TUR) requires that $\sigma \tau \geq 2 k_B \langle X_\tau \rangle^2/\text{Var} \, X_\tau$, where $X_\tau$ is the stochastic variable counting the number of full rotations in time $\tau$~\cite{Seifert_2016}.  This bound can be saturated, for instance consider a periodic interval, $S^1$, where the system evolves through a Langevin equation,
\be
\d Y_t = F \d t + \sqrt{2D} \d W_t,
\ee
for $F$ and $D$ constant. In this case, $Y_\tau  \sim Y_0 +  F \tau + \sqrt{2D} \mathcal{N}(0,\tau)$, so in the large $\tau$ limit, the average number of rotations is $\langle X_\tau \rangle \approx F \tau$, whereas the variance of this will be $\text{Var}\, X_\tau \approx 2D \tau$. Then $2 k_B \langle X_\tau \rangle^2/\text{Var} \, X_\tau \approx 2k_B (F \tau)^2 / 2 D \tau = k_B F^2  \tau / D = \sigma \tau$, since the entropy production rate for such a continuous system is $k_B F^2/D$~\cite{Seifert_2012}. \rev{Given any continuous Langevin system, one can construct a discrete Markov system which approximates the dynamics of the continuous system arbitrarily well and has an entropy production rate arbitrarily close to that of the continuous system~\cite{Skinner2021,Chernyak_2006}. Hence we can construct a discrete system with precision and entropy production arbitrarily close to the continuous system which saturates the TUR bound. However, whilst such a system will also saturate the TUR bound, it will be no more precise than an equilibrium system.} To see this with $A$ defined as some subset of $S^1$, note that the Langevin system at small times is dominated by diffusive behavior, with short time trajectories approximating Brownian motion. For a trajectory which starts in $A$ and ends in $B$, it will cross the boundary infinitely many times on short time scales before eventually leaving for good, meaning the waiting time statistics are dominated by this short term diffusive behavior, which could be generated by an equilibrium system.

We therefore cannot use directly the TUR bound to understand $\Gamma(\sigma)$ in the large $\sigma$ limit. Previously we have minimized over all unknown discrete network topologies with $N$ unknown states, and for large values of $\theta$, or small values of $\sigma$, the minimum is accessible for values of $N \leq 20$. However the asymptotic regime, $\theta \to 1$ \rev{would require values of $N$ that are numerically infeasible}. Instead, we formally take $N\to\infty$, and suppose that the region $A$ can be described through continuous coordinates undergoing Langevin dynamics. \rev{Importantly, such a coarse graining procedure, in which the limit of infinitely many discrete states is replaced by a continuous Langevin equation, does not necessarily preserve the entropy production rate~\cite{Busiello_2019,Herpich2020,Horowitz2015}, and hence it is not known if minimizing over one-dimensional Langevin system approximates the curve $\theta = \Gamma(\sigma)$ in the $\sigma \to \infty$ limit. However, we are motivated by the fact that the one-dimensional Langevin system saturates TUR for the stochastic clock, which counts by recording the number of full rotations, and for the stochastic timer, the optimized discrete systems appear Langevin-like in the large $\sigma$ limit. In any case, minimizing over the Langevin system provides an upper bound for the most optimized finite state Markov process, and in particular will allow us to show that we can achieve the performance of the TUR saturating stochastic clock to leading order in the large $\sigma$ limit with a stochastic timer.}

Formally, we take $B$ to be a single point, with $p^B = 1/2$, and $A$ to be $S^1$ with
\be
p^B + \int_0^1 \d x \; p(x) = 1,
\ee
as the total probability. Whilst in $B$, the system jumps to $x \in A$ at rate $a(x)\d x$, and whilst
at $x\in A$, the system jumps to $B$ at rate $b(x)$. Thus,
\be
\int \d x \; p(x) b(x) = \frac{1}{2}\int\d x \; a(x) = \frac{1}{2}.
\ee
Whilst in $A$, we assume trajectories $X_t$, follow the stochastic It\^o equation,
\be
\d X_t = F \d t + \sqrt{2D} \d W_t,
\ee
where $W_t$ is Brownian motion, and the decision to jump to $B$ or stay in $A$ is made at the start of
the time step, i.e. at $b(X_t)$, not $b(X_{t + \d t})$. We assume $F$ and $D$ are constants which greatly simplifies the analysis. To calculate the entropy production rate of this
system, we use that the entropy production rate is related to the relative probability of observing forward and reverse trajectories~\cite{Chernyak_2006},
\be
\sigma = k_B \lim_{\tau \to \infty} \frac{1}{\tau} \int_{0}^\tau \log \left[ \frac{ \mathbb{P}(\{x \}) }{ 
\mathbb{P} (\{ \tilde{x} \}) } \right]  \mathbb{P}(\{ x \}) \d \{x\},
\ee
where $\mathbb{P}(\{x\})$ is the probability of seeing some trajectory $\{x\}$ in $[0,\tau]$, with
$\{ \tilde{x} \}$ its time reversed counterpart. To evaluate this, we discretize time whilst in $A$,
so $X_t$ becomes $X_0,X_1,\dots X_i$ with a time step $\Delta t$, so $X_{i+1} = X_i + F \Delta t  + \sqrt{2D} \Delta W_i$, given that no jump to $B$ occured, with $\Delta W_{i}\sim \mathcal{N}(0,\Delta t)$. Then 
\be 
\mathbb{P}(X_{i+1}|X_i) = \frac{1-b(X_i)\Delta t}{\sqrt{ 2 \pi \Delta t}} 
\exp \left[ - \frac{ (\Delta X_i - F \Delta t)^2}{
4 D \Delta t} \right], 
\ee
where $\Delta X_i = X_{i+1} - X_i$. We have therefore that 
\begin{align}\nonumber
\log \frac{\mathbb{P}(X_{i+1}|X_i)}{\mathbb{P}(X_{i}|X_{i+1})} &= \log \frac{1 - b(X_i) \Delta t}{1 - 
b(X_{i+1})\Delta t} + \frac{1}{4 \Delta t} \left[ \frac{ (\Delta X_{i} + F \Delta t)^2}{D}
-\frac{ (-\Delta X_{i} + F \Delta t)^2}{D} \right] \\ 
&= \frac{F}{D} \Delta X_i  + O(\Delta t^{3/2}) = \frac{F^2}{D} \Delta t + \frac{F}{D}\sqrt{2D} \Delta W_i + O(\Delta t^{3/2})
\end{align}
in the limit $\Delta t \to 0$, and we have neglected higher order terms. In particular, for a continuous trajectory
$\{x\}$ that stays entirely within $A$, 
\begin{align}
\log \left[ \frac{ \mathbb{P}(\{x \}) }{ \mathbb{P} (\{ \tilde{x} \}) } \right] &= 
\log \left[ \frac{ \mathbb{P}(x(\tau)) }{ \mathbb{P} (x(0)) } \right]  + \frac{F^2 \tau}{D}   + F \sqrt{2/D} \int_0^\tau \d W_t.
\end{align}
and so for a trajectory $\{x\}$ that starts in $B$, jumps to $A$ at time $t=0$ and returns to $B$ at time $t=\tau$, 
\begin{align}
\log \left[ \frac{ \mathbb{P}(\{x \}) }{ \mathbb{P} (\{ \tilde{x} \}) } \right] &= 
\log \left[ \frac{ a(x(0)) b(x(\tau)) }{ a(x(\tau)) b(x(0))} \right]  + \frac{F^2 \tau}{D}   + F \sqrt{2/D} \int_0^\tau \d W_t.
\end{align}
Therefore, every jump from $B$ to $x \in A$ picks up an entropic contribution of $\log a(x)/b(x)$, the reverse jump
picks up a contribution of $\log b(x)/a(x)$, and while in $A$ entropy increases at a rate $F^2/D$, with the contribution from the integral $\int_0^\tau \d W_t$ being zero on average.
Taking the long time limit, as the probability distribution tends to the \rev{steady state distribution} $\{p(x),p^B\}$, the average rate
of entropy production becomes
\begin{align}
\sigma &= \frac{F^2}{2D}  + \int \d x \left[ p(x) b(x) \log (b(x)/a(x))  + p^B a(x) \log(a(x)/b(x)) \right].
\end{align}
In addition to the normalization conditions, the steady state probability distribution obeys the Fokker-Planck
equation,
\be
0 = -\partial_x (F p) + \partial_x^2 (D p) + p^B a - pb.
\ee
All that remains to formalize the minimization problem, is to calculate expressions for the first and second 
moments for the distribution of time spent in $A$. Since the rate at which the process leaves $A$ is $b(x)$,
the probability that a specific trajectory is still in $A$ at time $t$ is $\exp ( - \int_0^t b(X_s) \d s)$,
and therefore, if $v(x,t)$ is the probability of still being in $A$ at time $t$, given that at $t=0$, the
system was in state $x\in A$, then
\be
v(x,t) = \mathbb{E} \left[ e^{-\int_0^t b(X_s) \d s} | X_0 = x \right].
\ee
The Feynman-Kac formula allows us to instead solve the following PDE to calculate $v(x,t)$~\rev{\cite{pavliotis2014stochastic}},
\begin{align}
\partial_t v &= F \partial_x v + D \partial_x^2 v - bv, \\ \nonumber
v(x,0) &= 1,
\end{align}
\rev{and therefore, we can find the full wait time distribution given the initial condition by solving a PDE. However, to only calculate
the first and second moments, we only need to solve ODEs~\cite{pavliotis2014stochastic}.} Let $T^{(1)}(x)$ be the average wait time in $A$,
given the initial condition $x$, and $T^{(2)}(x)$ be the second moment. Now
\be 
T^{(1)}(x) = \int_0^{\infty} t (-\partial_t v) \d t = \int_0^{\infty} v \d t,
\ee
but if $\mathcal{L} = F \partial_x + D \partial_x^2 - b$, acts only on $x$, then
\be
\mathcal{L}T^{(1)}(x) = \int_0^{\infty} \mathcal{L}v \d t = \int_0^{\infty} \partial_t v \d t = -1.
\ee
Similarly, as
\be
T^{(2)} = \int_0^{\infty} t^2 (-\partial_t v) \d t = 2 \int_0^{\infty} t v \d t,
\ee
we have that
\be
\mathcal{L} T^{(2)} =  2 \int_0^{\infty} t \partial_t v \d t = -2 T^{(1)}.
\ee
We therefore need only solve two second order differential equations to recover the waiting time moments given
by
\begin{align} 
\langle t \rangle_A = \int T^{(1)} a \, \d x, \qquad
\langle t^2 \rangle_A = \int T^{(2)} a \, \d x.
\end{align}
The full minimization problem can therefore be described as 
\begin{subequations}
\begin{align}
\min \ \sigma  &= \sigma\left[ a,b,F,D \right] \text{ subject to} \\
\langle t^2 \rangle_A &= \int T^{(2)} a \d x, \\
\mathcal{L}^2 T^{(2)} &= 2, \\
\int a \d x &= 1, \\
\int p \d x &= 1/2, \\
\mathcal{L}^* p + \frac{1}{2} a &= 0,
\end{align}
\end{subequations}
where $\mathcal{L}^*f = -\partial_x(Ff) + \partial_x^2 (Df) - bf$, and the condition of $\langle t \rangle_A = 1$,
is enforced implicitly by the constraints on $a$ and $p$.
\subsection{Numerical minimization}
In order to minimize the constrained optimization problem over the infinite dimensional function space $C(S^1)$, 
we instead minimize over a finite dimensional Fourier basis, so that
\be
a(x) = \sum_{k = -N/2}^{N/2} a_k e^{2 \pi i k x},
\ee
and similar for $b,F,D$, where now we minimize over the $N/2 +1$ complex coefficients, $a_0,\dots, a_{N/2}$,
with $a_{-k} = a_k^*$. Taking a Fourier-Galerkin approach to solving the differential equations, the
solution $g(x)$ to $\mathcal{L}g =f$ is approximated as the $g$ for which 
\be 
\langle \mathcal{L}g - f, e^{2\pi i k} \rangle = 0
\ee
for $-N/2 \leq k \leq N/2$. Specifically, 
\begin{align} 
\mathcal{L}g - f &= \sum_{k,k'} (F_{k'} (2\pi i k) + D_{k'} (2\pi i k)^2 - b_{k'}) g_k e^{2\pi i (k + k') x}  - \sum_k f_k e^{2\pi i k} ,
\end{align}
so we have that 
\be
\sum_{k=-N/2}^{N/2} \mathcal{L}_{rk} g_k = f_r,
\ee
where
\be
\mathcal{L}_{rk} = (2\pi i k) F_{r-k} - D_{r-k} (2 \pi k)^2 - b_{r-k},
\ee
and terms $f_{r-k}$ for $|r-k| > N/2$ are taken as zero. Since the operator $\mathcal{L}^*$ is the adjoint of 
$\mathcal{L}$, solving the equation for $p$ uses the Hermitian adjoint of $\mathcal{L}_{rk}$, as
\be
\sum_k (\mathcal{L}^{\dagger})_{rk} p_k = -\frac{1}{2} a_r.
\ee
The minimization problem can now be stated as 
\begin{subequations}
\begin{align}
\min \sigma &= \sigma(a_k,b_k,F_k,D_k) \ \text{subject to} \\
\langle t^2 \rangle_A &= \sum_k T^{(2)}_k a_{-k} \\
T^{(2)}_r &= 2 \sum_k (\mathcal{L}^{-2})_{rk} \delta_{k0} \\
a_0 &= 1 \\
p_0 &= \frac{1}{2} \\
p_r &=  -\frac{1}{2} \sum_k (\mathcal{L}^{-\dagger})_{rk} a_k
\end{align}
\end{subequations}
No explicit expression for $\sigma$ in terms of the coefficients is possible due 
to the non-linear terms, but we can evaluate it by numerical
integration, which converges exponentially fast for periodic functions.
The condition $a,b,D \geq 0$ is not enforced directly, but through a series of trial points $x_i$, so
$a(x_i) \geq 0$, which is a linear constraint on the coefficients $a_k$. In addition, we take only
the first $M/2$ coefficients of $a,b$ to be non-zero, and for now take $F$ and $D$ to be constant. This
ensures that we solve the problem accurately for given $a,b$, and regularizes the solution.
\subsection{Analytic gradients}
We calculate the following gradients analytically to perform the minimization,
\begin{subequations}
\begin{align}
\frac{\partial \sigma} {\partial F} &= F/D + \int \d x\ \frac{\partial p}{\partial F} b\log(b/a)\\
\frac{\partial \sigma} {\partial D} &= -F^2/ 2D^2  + \int \d x\ \frac{\partial p}{\partial D} b\log(b/a)\\
\frac{\partial \sigma} {\partial b_0} &= \int \d x \left[ \frac{\partial p}{\partial b_0} b \log(b/a) + p (\log (b/a)+1)  
- p^B a / b \right],
\end{align}
\end{subequations}
where $\partial p /\partial b_0$ can be computed from 
\be
\frac{\partial p_k}{\partial b_0}  = \sum_r (\mathcal{L}^{-\dagger})_{kr} p_r,
\ee
and
\begin{subequations}
\begin{align}
\frac{\partial p_k}{\partial F} &= \sum_j (\mathcal{L}^{-\dagger})_{kj} (2\pi i j) p_j, \\
\frac{\partial p_k}{\partial D} &= \sum_j (\mathcal{L}^{-\dagger})_{kj}(2 \pi j)^2 p_j,
\end{align}
\end{subequations}

The rest are more involved as we must differentiate with respect to both the real and imaginary parts $a_k^r$, and
$a_k^i$ respectively.  We first note that
\begin{subequations}
\begin{align}
\frac{\partial a}{\partial a_k^r} &= 2 \cos ( 2 \pi k x) \\ 
\frac{\partial a}{\partial a_k^i} &= -2 \sin ( 2 \pi k x) \\
\frac{\partial \sigma}{\partial a_k^r} &= \int \d x \left[ \frac{\partial p}{\partial a_k^r} b \log (b/a) 
-\frac{2pb\cos{2\pi k x}}{a}  + 2p^B \cos (2 \pi k x) ( \log (a/b) + 1)\right] \\
\frac{\partial \sigma}{\partial a_k^i} &= \int \d x \left[ \frac{\partial p}{\partial a_k^i} b \log (b/a) 
+\frac{2pb\sin{2\pi k x}}{a} - 2p^B \sin(2 \pi k x) ( \log (a/b) + 1)\right],
\end{align}
\end{subequations}

where
\begin{subequations}
\begin{align}
\frac{\partial p_k}{\partial a^r_j} &= -\frac{1}{2} [(\mathcal{L}^{-\dagger})_{kj} 
+ (\mathcal{L}^{-\dagger})_{k,-j}], \\
\frac{\partial p_k}{\partial a^i_j} &= -\frac{i}{2} [(\mathcal{L}^{-\dagger})_{kj}
- (\mathcal{L}^{-\dagger})_{k, - j}],
\end{align}
\end{subequations}
Next,
\begin{subequations}
\begin{align}
\frac{\partial \sigma}{\partial b_k^r} &= \int \d x \left[ \frac{\partial p}{\partial b_k^r} b \log (b/a) 
+2p\cos(2\pi k x)( \log(b/a) + 1)  - 2p^B a \cos (2 \pi k x) / b \right] \\
\frac{\partial \sigma}{\partial b_k^i} &= \int \d x \left[ \frac{\partial p}{\partial b_k^i} b \log (b/a) 
-2p\sin(2\pi k x)(\log(b/a)+1)  + 2p^B a \sin(2 \pi k x)/b \right],
\end{align}
\end{subequations}
where
\begin{subequations}
\begin{align}
\frac{\partial p_k}{\partial b^r_j} &=  \sum_l (\mathcal{L}^{-\dagger})_{kl} (p_{l-j} + p_{j+l}), \\
\frac{\partial p_k}{\partial b^i_j} &= i \sum_l (\mathcal{L}^{-\dagger})_{kl}(p_{l-j} - p_{j+l}).
\end{align}
\end{subequations}
Now, defining
\begin{subequations} 
\begin{align}
P_l &= \int \d x e^{2 \pi i l x} b \log (b/a), \\
Q_l &= \int \d x e^{2 \pi i l x} [- 2pb/a + 2p^B(\log (a/b) + 1) ], \\
R_l &= \int \d x e^{2 \pi i l x} [ 2p(\log(b/a) + 1) - 2 p^B a/b ],
\end{align}
\end{subequations}
we can write the derivatives as 
\begin{subequations}
\begin{align}
\frac{\partial \sigma} {\partial F} &= F/D + \sum_l P_l \frac{\partial p_l}{\partial F} \\
\frac{\partial \sigma} {\partial D} &= -F^2/ 2D^2  + \sum_l P_l \frac{\partial p_l}{\partial D} \\
\frac{\partial \sigma} {\partial a_k^r} &= \sum_l P_l \frac{\partial p_l}{\partial a_k^r} +  \text{Re}(Q_k) \\
\frac{\partial \sigma} {\partial a_k^i} &= \sum_l P_l \frac{\partial p_l}{\partial a_k^i} - \text{Im}(Q_k) \\
\frac{\partial \sigma} {\partial b_0} &= \sum_l P_l \frac{\partial p_l}{\partial b_0} + \frac{1}{2} R_0 \\
\frac{\partial \sigma} {\partial b_k^r} &= \sum_l P_l \frac{\partial p_l}{\partial b_k^r} +  \text{Re}(R_k) \\
\frac{\partial \sigma} {\partial b_k^i} &= \sum_l P_l \frac{\partial p_l}{\partial b_k^i} - \text{Im}(R_k).
\end{align}
\end{subequations}
The only integrals now occur within the terms $P_l,Q_l,R_l$, and can be evaluated by using a fast fourier transform. If we further
identify $w_j = [\sum_l \mathcal{L}^{-1}_{jl} P_l^* ]^*$, then 
\begin{subequations}
\begin{align}
\frac{\partial \sigma} {\partial F} &= F/D + i\sum_j 2\pi  j w_j \\
\frac{\partial \sigma} {\partial D} &= -F^2/ 2D^2  + \sum_j (2\pi j)^2 w_j  \\
\frac{\partial \sigma} {\partial a_k^r} &= -\frac{1}{2}(w_k + w_{-k}) +  \text{Re}(Q_k) \\
\frac{\partial \sigma} {\partial a_k^i} &= -\frac{i}{2}(w_k - w_{-k}) - \text{Im}(Q_k) \\
\frac{\partial \sigma} {\partial b_0} &= \sum_l w_l p_l + \frac{1}{2} R_0 \\
\frac{\partial \sigma} {\partial b_k^r} &= \sum_l w_l (p_{l-k} + p_{l+k})  +  \text{Re}(R_k) \\
\frac{\partial \sigma} {\partial b_k^i} &= i \sum_l w_l (p_{l-k} - p_{l+k}) - \text{Im}(R_k).
\end{align}
\end{subequations}
We must also differentiate the constraints, the first of which can be expressed as 
$c_1 = \sum_{k} T^{(2)}_k a_{-k} - \langle t^2 \rangle_A$, so that $c_1=0$. We have immediately that
\begin{subequations}
\begin{align}
\frac{\partial c_1} {\partial F} &= \sum_l a_{-l} \frac{\partial T^{(2)}_l}{\partial F} \\
\frac{\partial c_1} {\partial D} &= \sum_l a_{-l} \frac{\partial T^{(2)}_l}{\partial D} \\
\frac{\partial c_1} {\partial a_k^r} &= T^{(2)}_k + T^{(2)}_{-k}\\
\frac{\partial c_1} {\partial a_k^i} &= -i T^{(2)}_k + i T^{(2)}_{-k}\\
\frac{\partial c_1} {\partial b_0} &= \sum_l a_{-l} \frac{\partial T^{(2)}_l}{\partial b_0}  \\
\frac{\partial c_1} {\partial b_k^r} &= \sum_l a_{-l} \frac{\partial T^{(2)}_l}{\partial b_k^r}  \\
\frac{\partial c_1} {\partial b_k^i} &= \sum_l a_{-l} \frac{\partial T^{(2)}_l}{\partial b_k^i},
\end{align}
\end{subequations}
and
\begin{subequations}
\begin{align}
\sum_{k,l}\mathcal{L}_{jk}\mathcal{L}_{kl} \frac{\partial T^{(2)}_l}{\partial F} &= -i \sum_l \left[ (2\pi j) 
\mathcal{L}_{jl}T^{(2)}_l + \mathcal{L}_{jl} (2\pi l) T^{(2)}_l \right] \\
\sum_{k,l}\mathcal{L}_{jk}\mathcal{L}_{kl} \frac{\partial T^{(2)}_l}{\partial D} &=  \sum_l \left[ (2\pi j)^2 
\mathcal{L}_{jl}T^{(2)}_l + \mathcal{L}_{jl} (2\pi l)^2 T^{(2)}_l \right] \\
\sum_{k,l}\mathcal{L}_{jk}\mathcal{L}_{kl} \frac{\partial T^{(2)}_l}{\partial b_s^r} &=  \sum_l \left[ 
\mathcal{L}_{j-s,l} + \mathcal{L}_{j+s,l} + \mathcal{L}_{j,l + s} + \mathcal{L}_{j,l-s} \right] T^{(2)}_l  \\
\sum_{k,l}\mathcal{L}_{jk}\mathcal{L}_{kl} \frac{\partial T^{(2)}_l}{\partial b_s^i} &=  i\sum_l \left[ 
\mathcal{L}_{j-s,l} - \mathcal{L}_{j+s,l} + \mathcal{L}_{j,l + s} - \mathcal{L}_{j,l-s} \right] T^{(2)}_l 
\end{align}
\end{subequations}
The common form is $\partial c/\partial y = \sum_l a_{-l} f_l$, where $f$ is some vector solving 
$ \mathcal{L}^2 f = g$. There are many such $y$ however, and we wish to solve as few linear equations as possible.
However, we can rewrite as $\partial c / \partial y = \sum_{k,l} [(\mathcal{L}^2)_{kl}^{-\dagger}a_l]^* g_k 
= \sum_k v_{k}^* g_k$, where $v_k = (\mathcal{L}^2)_{kl}^{-\dagger}a_l$. We therefore have that
\begin{subequations}
\begin{align}
 \frac{\partial c_1 }{\partial F} &= -i \sum_{l,j} v_j^* \left[ (2\pi j) 
\mathcal{L}_{jl}T^{(2)}_l + \mathcal{L}_{jl} (2\pi l) T^{(2)}_l \right] \\
 \frac{\partial c_1 }{\partial D} &=  \sum_{l,j} v_j^* \left[ (2\pi j)^2 
\mathcal{L}_{jl}T^{(2)}_l + \mathcal{L}_{jl} (2\pi l)^2 T^{(2)}_l \right] \\
 \frac{\partial c_1}{\partial b_s^r} &=  \sum_{l,j} v_j^* \left[ 
\mathcal{L}_{j-s,l} + \mathcal{L}_{j+s,l} + \mathcal{L}_{j,l + s} + \mathcal{L}_{j,l-s} \right] T^{(2)}_l  \\
 \frac{\partial c_1 }{\partial b_s^i} &=  i\sum_{l,j} v_j^* \left[ 
\mathcal{L}_{j-s,l} - \mathcal{L}_{j+s,l} + \mathcal{L}_{j,l + s} - \mathcal{L}_{j,l-s} \right] T^{(2)}_l 
\end{align}
\end{subequations}
We can simplify further by defining $u_j = \sum_l \mathcal{L}_{jl} T^{(2)}_l$, and recall that $-2 p_j = 
\sum_l \mathcal{L}^{-\dagger}_{jl} a_l$, so that 
\begin{subequations}
\begin{align}
 \frac{\partial c_1 }{\partial F} &= -2 \pi i \sum_{j} j[v_j^* u_j -2 p_j^*  T_j^{(2)}] \\
 \frac{\partial c_1 }{\partial D} &=  (2\pi)^2 \sum_{j} j^2 [v_j^* u_j -2 p_j^* T_j^{(2)}] \\
 \frac{\partial c_1}{\partial b_s^r} &=  \sum_{j} v_j^* ( u_{j-s} + u_{j+s}) - 2(p_{l+s}^* + p_{l-s}^*) T^{(2)}_l \\
 \frac{\partial c_1 }{\partial b_s^i} &=  i\sum_{l,j} v_j^* ( u_{j-s} - u_{j+s})- 2(p_{l+s}^* - p_{l-s}^*) T^{(2)}_l 
\end{align}
\end{subequations}

\subsection{Asymptotic limit}
Using the continuous Langevin equations, we can reach larger values of $\sigma$ in our numerical minimization than we could by working with the finite state Markov system. However, we still seek an asymptotic formula that will allow us to extend to arbitrarily large values of $\sigma$ beyond what is numerically possible. 

In the formal limit of infinite precision, a system would have $D=0$,  $F>0$, together with $a$ and $b$ as $\delta$-functions, so that a trajectory would enter $A$ at a point specified by $a$, travel deterministically with velocity $F$, before leaving at the point specified by $b$ after spending exactly $t=1$ in $A$. Based on intuition from this infinite precision system, we will derive an asymptotic limit as $\text{Var}\,t_A \to 0$. Specifically, for analytic tractability we take $a(x) = a_0 \delta(x_0) + a_1\delta(x_1)$, $b(x) = b_0 \delta(x_0) + b_1 \delta(x_1)$,  allowing us to minimize over a system specified by a finite number of variables rather than arbitrary functions $a(x)$, $b(x)$. Whilst we expect, and observe numerically, the functions $a(x)$ and $b(x)$ to be sharply peaked in the limit, initially assuming $\delta$-functions rather than finite width Gaussians or arbitrary peaked functions may affect terms in the asymptotic expansion. However, after we perform this minimization, we find $a$ and $b$ do not show up at leading order, and hence we expect that the effect of changing from a $\delta$-function to a sharply peaked Gaussian will show up at terms smaller than the largest terms involving $a$ and $b$, which is higher than the leading two term asymptotic expansion derived here.

\begin{figure}
\includegraphics{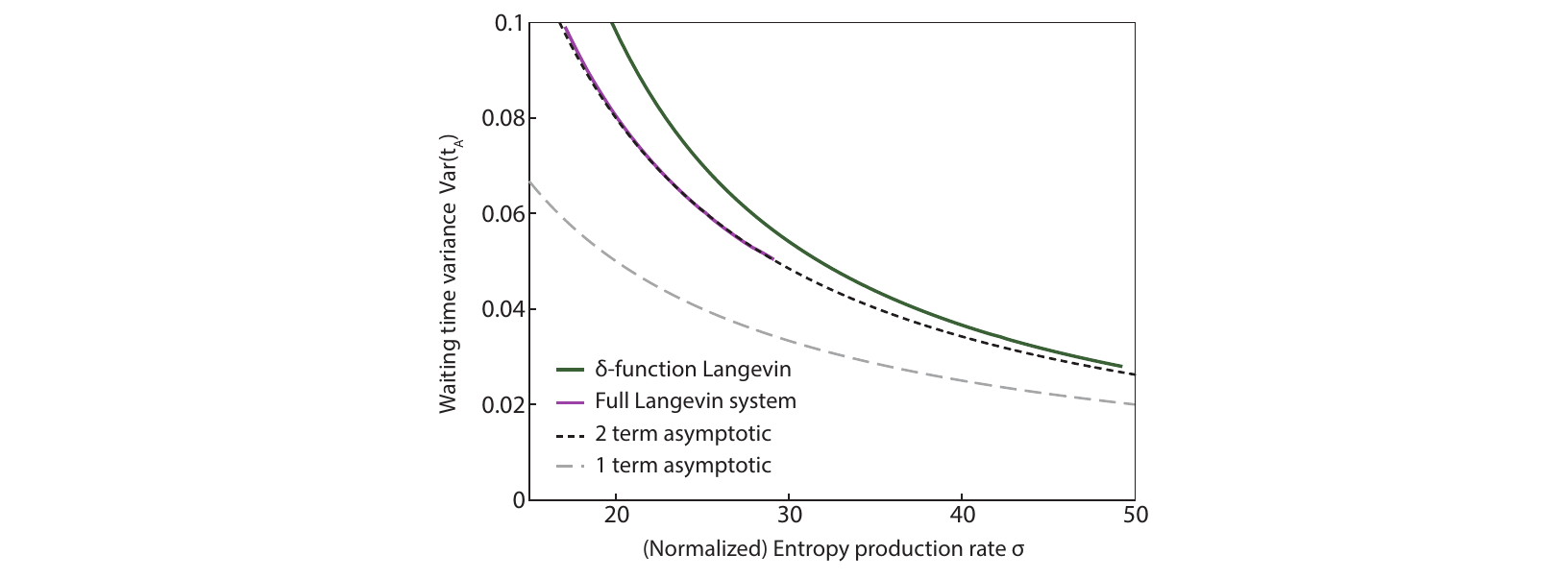}
\caption{\label{fig:LargeAsymp} Comparison of asymptotic bounds and numerical minimization for the continuous Langevin system. The one term, $\text{Var}\, t_A \sim 1/ \sigma$ (grey), and two term $\text{Var}\, t_A \sim 1/ \sigma + 4 \log \sigma / \sigma^2$ (black), asymptotic results are shown. Minimizing over the full Langevin system (purple), with $a$, $b$ arbitrary functions, shows the two term asymptotic result is achievable, although computational constraints prevent extending this curve for larger $\sigma$. Performing a numerical minimization with $a$, $b$ as $\delta$-functions, as was assumed to derive the asymptotics, also agrees with the asymptotics in the limit, but takes 
longer to converge (green).}
\end{figure}

The problem is then as follows, 
\begin{subequations}
\begin{align}
\min \sigma[D,F,a_0,a_1,b_0,b_1] &=  \frac{F^2}{2D} + \sum_{i=1,2}\; p(x_i) b_i \log(b_i/a_i) + \frac{1}{2} a_i \log(a_i/b_i)\ \text{subject to}, \\
\langle t^2 \rangle_A &= a_0 T^{(2)}(x_0) + a_1 T^{(2)}(x_1) \\
\mathcal{L}^2 T^{(2)} &= 2 \\
a_0 + a_1 &= 1 \\
\int p\, \d x &= 1 \\
\mathcal{L}^* p + \frac{1}{2}a &=0,
\end{align}
\end{subequations}
where the presence of $\delta$-functions leads to jump conditions for the differential equations. For instance the  equation for $p$ becomes
$D \partial_x^2 p - F \partial_x p = 0$, away from $x_i$, where $p$ is everywhere continuous with discontinuous first derivative satisfying the jump conditions $D [\partial_x p]^{x_i^+}_{x_i^-}  - b_i p(x_i) + \frac{1}{2} a_i = 0$ for $i=1,2$. Solving these equations to find $p$ is possible analytically, and moreover it is possible to eliminate $a_0$, $a_1$ from the equations, but the resulting expressions are complex. We therefore provide a Mathematica notebook\footnote{\texttt{\href{https://github.com/Dom-Skinner/EntropyProductionFromWaitingTimes}{https://github.com/Dom-Skinner/EntropyProductionFromWaitingTimes}}}, which contains explicit expressions for $p$, $a_0$, $a_1$, and hence $\sigma$ and $\langle t^2 \rangle_A$ in terms of the variables $b_i,F,D,x_i$. Taking the limit $D\to0$, and exploring how the remaining parameters scale, we find the asymptotic relation
\be
\langle t^2 \rangle_A  \sim 1 + \frac{1}{\sigma} + \frac{4 \log \sigma}{\sigma^2},
\ee
provides the most precision for a given $\sigma$ in the $\sigma\to\infty$ limit. We see that the numerical minimization agrees well with this asymptotic formula for values of $\sigma$ as low as $20$, Fig.~\ref{fig:LargeAsymp}. We can interpret the resulting system as essentially the continuous Langevin system which saturates the TUR bound, but with an additional cost of having to determine exactly when one rotation has occurred and subsequently changing to \rev{macro-state} $B$~\cite{Seifert_2016,Pearson2021}. However, this
additional cost is of order $O(\log \sigma / \sigma ^2 )$, and so is not leading order in the $\sigma \to \infty$ limit.
%%%%%%%%%%%%%%%%%%%%%%%%%%%%%%%%%%%%%%%%%%%%%%%%%%%%%%%%%%%%
\section{Further examples from the literature}
In addition to the examples investigated in the main text, there are a large number of examples in the literature which measure non-equilibrium waiting time distributions from which we can bound the entropy production rate. We include a number of examples in Table~\ref{tab:Ex}, where we have computed $\sigma_T$ directly from experimental waiting time distribution histograms. Where both states of the two \rev{macro-state} system have non-equilibrium waiting time distributions we selected $A$ as the one to give the largest $\sigma_T$ bound, and where $\langle t \rangle_B$ is not explicitly given we use an order of magnitude estimate.

\begin{table}\caption{\label{tab:Ex}Further examples and $\sigma_T$ bound}
\begin{center}
\begin{tabularx}{\textwidth}{p{5cm}@{\hskip 0.3cm} p{1.7cm} p{1.7cm} p{1.2cm} p{2.2cm}@{\hskip 0.3cm} p{3.6cm} } 
System & $\langle t \rangle_A$ & $\langle t \rangle_B$ & $\frac{\langle t^2 \rangle_A }{\langle t \rangle_A ^2}$ & $\sigma_T$ & Reference \\ [1ex] 
\hline\hline & \\[-2ex]
\multicolumn{1}{l}{Swim-turn dynamics} \\[1ex] 
\emph{Pseudomonas putida} & 1.5s &  0.13s & 1.4 & 4.6 $k_B $/s & Fig. 4(b) in Ref.~\cite{THEVES2013} \\ [1ex] 
\emph{Chlamydomonas} & 2.0s &  11.2s & 1.2 & 1.4 $k_B $/s & Fig. 4(d) in Ref.~\cite{Polin2009} \\ [1ex] 
\hline& \\[-2ex]
\multicolumn{1}{l}{Swarm reversal dynamics} \\ [1ex] 
\emph{Myxococcus xanthus} & 520s &  $\approx 0$s & 1.2 & 0.035 $k_B $/s & Fig. S1 in Ref.~\cite{Wu2009} \\ [1ex] 
\emph{Myxococcus xanthus} collective waves & 310s &  $\approx 0$s & 1.3 & 0.042 $k_B $/s & Fig. 2(a) in Ref.~\cite{Sliusarenko2006} \\ [1ex] 
\emph{Paenibacillus dendritiformis} & 20s &  $\approx 4$s & 1.3 & 0.43 $k_B $/s & Fig. 3(b) in Ref.~\cite{Beer2013} \\ [1ex] 
\hline& \\[-2ex]
Migrating breast cancer cell repolarization time & $8.3\times 10^3$s &  $\approx 10^4$s  & 1.4 & $5\times10^{-4} k_B $/s & Fig. 4(b) (inset L170) in Ref.~\cite{Zhou2020} \\ [1ex] 
\hline& \\[-2ex]
Visual perception switching & 4.7s &  4.7s  & 1.5 & $0.78 k_B $/s & Fig. 3(a)(i) in Ref.~\cite{Krug2008} \\ [1ex] 
\hline& \\[-2ex]
\multicolumn{1}{l}{Animal and insect flight duration} \\ [1ex] 
\emph{Uria lomvia} & 800s & $\approx$ 500s & 1.7 & $2.6\times10^{-3}k_B$/s & Fig. 4 in Ref~\cite{Elliott2009}. \\[1ex]
\emph{Bombus terrestris} & 2.5s & $\approx$ 5s & 1.7 & $0.38 k_B$/s & Fig. 5d in Ref~\cite{raine2007}. \\[1ex]
\hline& \\[-2ex]
\emph{Escherichia coli} flagellar rotation direction switching & 0.020s & 0.019s & 1.9 & 33 $k_B$/s & Fig. 4(a, inset middle) in Ref.~\cite{BerryScience} \\[1ex]
\hline & \\[-2ex]
\emph{Bacillus subtilis} gene state switching & $1.2\times 10^4$s & $9.5\times 10^4$s & 1.1 & $3.0 \times 10^{-4}k_B$/s & Fig. 2(d, f) in Ref.~\cite{Norman2013} \\[1ex]
\hline& \\[-2ex]
Binding-unbinding times of kinesin-1 molecular motor & $0.075$s & $0.019$s & 1.3 & $120 k_B$/s & Fig. 2(d) in Ref.~\cite{Isojima2016} \\[1ex]
\hline& \\[-2ex]
RNA polymerase cluster lifetime & $5.1$s & $\approx 500$s & 1.6 & $9.8\times 10^{-3} k_B$/s & Fig. 2(e) in Ref.~\cite{Cisse664}
\end{tabularx}
\end{center}
\end{table}